\documentclass[twocolumn,final]{svjour3}
\smartqed
\journalname{Theoret.~Chem.~Acc.}
\usepackage{graphicx}
\graphicspath{{pic/}}
\usepackage{subfigure}
\usepackage[utf8]{inputenc}
\usepackage{amsmath}
\usepackage{graphics}
\usepackage{epstopdf}
\usepackage{dcolumn}
\usepackage{bm}
\usepackage{rotating}
\usepackage{bbm}
\usepackage{verbatim}
\usepackage{txfonts}
\usepackage[cspex,bbgreekl]{mathbbol}
\usepackage{subfigure}
\usepackage{color}
\usepackage[version=3]{mhchem}
\usepackage[numbers,sort&compress]{natbib}
\usepackage{hyperref}
\hypersetup{
  colorlinks=true,
  linkcolor=blue,
  citecolor=red,
  linktoc=all,
  pdftitle=Local Random Phase Approximation with Projected Oscillator Orbitals,
  pdfdisplaydoctitle=true,
  pdfpagelayout=SinglePage,
  pdfstartview=Fit,
  pdfstartpage=1,
  bookmarksopen=false
}

\newcommand{\bra}[1]{\ensuremath{\langle #1 \vert\,  }}
\newcommand{\ket}[1]{\ensuremath{\,\vert   #1 \rangle}}
\newcommand{\braket}[2]{\ensuremath{\langle #1 \,\vert\,          #2 \rangle}}
\renewcommand{\b}[1]{\ensuremath{\mathbf{#1}}}
\newcommand\POO{{\text{POO}}}
\renewcommand\over[1]{{\overline{#1}}}

\newcommand{\Ra}{\mathcal{R}}

\usepackage[normalem]{ulem}
\usepackage{color,xcolor}

\begin{document}

\title{Local Random Phase Approximation with
Projected Oscillator Orbitals
\thanks{Dedicated to Prof.~P\'eter Surj\'an on the occasion
of his 60th birthday.}
}
\subtitle{}

\titlerunning{Local RPA with POOs}
\author{
     Bastien Mussard
\and J\'anos~G.~\'Angy\'an
}
\institute{
     Bastien Mussard \at
Sorbonne Universités, UPMC Univ Paris 06, Institut du Calcul et de la Simulation, F-75005, Paris, France
Sorbonne Universités, UPMC Univ Paris 06, UMR 7616, Laboratoire de Chimie Théorique, F-75005 Paris, France
CNRS, UMR 7616, Laboratoire de Chimie Théorique, F-75005 Paris, France
     \email{bastien.mussard@upmc.fr}
\and J\'anos~G.~\'Angy\'an \at
     Universit\'e de Lorraine, Institut Jean Barriol, CRM2, UMR 7036, Vandoeuvre-l\`es-Nancy, F-54506, FRANCE
 and CNRS, Institut Jean Barriol, CRM2, UMR 7036, Vandoeuvre-l\`es-Nancy, F-54506, FRANCE
     \email{janos.angyan@univ-lorraine.fr}
}
\authorrunning{B.~Mussard \& J.G.~\'Angy\'an}
\date{\today}

\maketitle

\begin{abstract}
An approximation to the many-body London dispersion energy in molecular  systems is expressed as a functional of the occupied orbitals only. The method is based on the local-RPA theory. The occupied orbitals are localized molecular orbitals and the virtual space is described by projected oscillator orbitals, \textit{i.e.}\ functions obtained by multiplying occupied localized orbitals with solid spherical harmonic polynomials having their origin at the orbital centroids. Since we are interested in the long-range part of the correlation energy, responsible for dispersion forces, the electron repulsion is approximated by its multipolar expansion. This procedure leads to a fully non-empirical long-range correlation energy expression. Molecular dispersion coefficients calculated from determinant wave functions obtained by a range-separated hybrid method reproduce experimental values with less than 15\% error.
\end{abstract}

\section{Introduction}

According to the Perdew's popular classification of density functional
approximations (DFA) \cite{Perdew:01f}, the random phase approximation (RPA) is situated on the
highest, fifth rung of Jacob's ladder, which leads from the simplest Hartree
level towards the ``heaven'' corresponding to the exact solution of the
Schrödinger equation. When stepping upwards on Jacob's ladder, one
uses more and more ingredients of the Kohn-Sham single determinant.
Starting from the lowest rungs, the density, its gradient, the full set of occupied orbitals are successively necessary for the construction of the functional.
At the highest rung
DFA is usually based on many-body methods, which require the knowledge of the complete set of occupied and virtual orbitals. Such methods have the drawback that the size of the virtual orbital space can be very large even in a moderately sized atomic orbital basis. In the case of plane wave calculations the virtual space  can become even prohibitively large. One solution to keep the size of matrices in reasonable limits  makes recourse to an auxiliary basis set to expand the occupied-virtual product functions. Such approaches are known in quantum chemistry as  resolution of identity \cite{Eichkorn:95} or density-fitting \cite{Whitten:73,Roeggen:08} methods. Similar advantages can be achieved by Cholesky decomposition \cite{Beebe:77} techniques. In plane wave calculations the plane wave basis itself can be used to expand the product states \cite{Harl:08}. Further gain can be achieved by projection methods, which avoid any explicit reference to  virtual orbitals. Such a technique, named projective dielectric eigenpotential (PDEP) method,  has been successfully applied to construct the dielectric function in plane wave RPA calculations \cite{Nguyen:09,Wilson:09,Wilson:08}. Recently, Rocca succeeded to reduce further the size of the problem by a prescreened set of virtuals \cite{Rocca:14}.

The purpose of  the present work is to demonstrate that  an \textit{approximate} variant of the RPA (and also of the MP2)  correlation energy can be expressed by using quantities that are computable from occupied orbitals alone. In this  context, one should mention the beautiful result, which has been obtained by Surj\'an, who has shown that the  MP2 correlation energy, exempt of any further approximation in a given basis set, can be reformulated as a functional of the Hartree-Fock density matrix, i.e.\ using exclusively  the occupied orbitals \cite{Surjan:05}. Our main interest is not to reproduce the full correlation energy with high numerical precision, but we focus our attention to a well-defined part of it, namely on the long-range dynamical correlation energy, which is usually taken responsible for the London dispersion forces.

It is now well-documented that most of the conventional density functional calculations in the Kohn-Sham framework, unless special corrections are added to the total energy, are unable to grasp the physics of these long-range forces.  Various fairly successful dispersion correction schemes are known, but although most of them were claimed to posses an essentially \textit{ab initio} character, they use, without exception, some ``external'' data, like atomic polarizabilities, atomic radii, \textit{etc}\dots \cite{Johnson:05,Johnson:09,Grimme:10,Grimme:11c,Tkatchenko:09,DiStasioJr:14, Reilly:15a}. Even if these quantities are usually not taken from experimental sources, and rather originate from ab initio computations, their presence deprives these theories of their self-contained character.
Hence, although we do not pretend that the rather drastic approximations which are going to be implemented in the following allow us to achieve results of a quality comparable to the precision attained by carefully fine-tuned methodologies, we argue that the ingredients of the present approach originate from a controlled series of approximations and do not use any ``external'' inputs. Moreover, in contrast to the relatively costly and sophisticated methods, like RPA, we do not need virtual orbitals, \textit{i.e.}\ we stay on the fourth rung of Jacob's ladder.

Our approach relies on the use of localized occupied molecular orbitals (LMOs) that can be obtained relatively easily by a unitary transformation in the subspace of occupied orbitals, according to either an external or an internal localization criterion \cite{Surjan:89b}. The localization of virtual orbitals is much more difficult, since the usual localization criteria for the occupied orbitals lead often to divergent results. It is to be noted that recently a significant progress has been reported \cite{Hoyvik:12} for the efficient localization of virtual orbitals. However, we follow another strategy here and build  excited determinants using localized functions which are able to span the essential part of the virtual space. One of the most popular local correlation approaches in this spirit  consists in using the atomic orbital (AO) basis functions to represent the virtual space. They are made orthogonal to the occupied space by projection, leading to the projected AOs (PAOs) techniques \cite{Pulay:83b,Pulay:86,Boughton:93}.  The locality of these functions is guaranteed by construction, even if it may be somewhat deteriorated by the projection procedure.

In the present work we are going to revisit and explore a quite old idea of Foster and Boys from the early sixties \cite{Foster:60a,Foster:60b,Boys:66}. The main concern of these authors was to construct a set of virtual orbitals directly from the set of occupied LMOs by multiplying them with solid spherical harmonic functions centered on the barycenter of the LMO. The orthogonality of these new functions can be ensured by a projection procedure.
Boys and Foster called these new functions, obtained after multiplication, \textit{oscillator orbitals} (OOs), and after removing the components of the OOs in the space of the occupied orbitals they may be called projected oscillator orbitals (POOs). Very few articles in the literature mention Boys' oscillator orbitals \cite{Alagona:72}, probably because it had no particular numerical advantage in high-precision configuration interaction calculations and its practical implementation raised a number of complications which could be avoided by more straightforward algorithms, like the use of the full set of virtual molecular orbitals (VMOs). We have found only a single,  very recent article, which  referred to the notion of oscillator orbitals \cite{Santolini:15} as a useful concept, but not as a practical computational tool. To the best of our knowledge, the mathematical  implications of using oscillator orbitals to define the virtual space has never been rigorously studied. Such an analysis is beyond the scope of the present study: it is going to be the subject of a forthcoming publication. 

The POOs are non-orthogonal among each other, which is at the origin one of the complications mentioned above. This problem can be handled just like in the case of the PAOs, therefore a theory of electron correlation based on POOs can follow a similar reasoning as local correlation methods using PAOs~ \cite{Knowles:00}. In particular, in this paper, the RPA method will be reformulated for a virtual solace constructed from POOs.

It is worthwhile to mention that the projected oscillator orbitals 
bear some similarities to the trial perturbed wave function in the 
variation-perturbational technique 
associated with the names of Kirkwood \cite{Kirkwood:32}, Pople and Schofield \cite{Pople:57} (KPS), 
to calculate multipole molecular polarizabilities. The closely related Karplus-Kolker \cite{Karplus:63a,Karplus:63b} (KK) method and its variants \cite{Rivail:78,Rivail:79} use a similar \textit{Ansatz} for the perturbed orbitals.  
In these latter methods, which were formulated originally as  simplified perturbed Hartree-Fock theories,  the first order perturbed 
wave function is a determinant with first order orbitals $\psi^{(1)}_i$ which are taken in the following product form \cite{Sadlej:71a} 
$\psi^{(1)}_i = g_i \psi_i - \sum_k  \braket{\psi_k}{g_i \psi_i}\psi_k $,
where $g_i$ are linear combination of some analytically defined functions, like polynomials. As we shall see, the principal difference of this \textit{Ansatz} and the POOs is that in the former case one multiplies the occupied canonical orbitals with the function $g_i$, while the oscillator orbitals are constructed from localized orbitals.

As mentioned previously, our main focus is the modeling of London dispersion forces. It has been demonstrated in our earlier works \cite{Toulouse:09,Zhu:10,Toulouse:11} that the essential physical ingredients of London dispersion forces are contained in the range-separated hybrid RPA method, where the short-range correlation effects are described within a DFA and the long-range exchange and correlation are handled at the long-range Hartree-Fock and long-range RPA levels, respectively. Among  numerous possible formulations of the RPA \cite{Angyan:11,Toulouse:11}, we have chosen to adopt  the variant based on the ring-diagram approximation to the coupled cluster doubles theory. The relevant amplitude equations will be rewritten with the help of POOs leading to simplified working equations, which do not refer to virtual orbitals explicitly. For the sake of comparison we are going to study the case where the POOs are expanded in the virtual orbital space. The long-range electron repulsion integrals, appearing in the range-separated correlation energy expression, can be reasonably approximated by a truncated multipole expansion. It means that in addition to the well-known improved convergence properties of the correlation energy with respect to the size of the basis set, one is able to control the convergence through the selection of the multipolar nature of the excitations, leading to a possibility of further computational gain. 

The exploitation of localized orbitals for dispersion energy calculations has already been proposed since the early works on local correlation methods \cite{Kapuy:91,Saebo:93,Hetzer:98,Usvyat:07,Chermak:12}. In classical and semiclassical models most often the atoms are selected as force centers; only a few works exploit the advantages related to the use of two-center localized orbitals and lone pairs. A notable exception is the recent work of Silvestrelli and coworkers \cite{Silvestrelli:09,Silvestrelli:09a,Ambrosetti:12,Silvestrelli:13a,Silvestrelli:14}, who adapted the Tkatchenko-Scheffler model \cite{Tkatchenko:09} for maximally localized Wannier functions, which are essentially  Boys' localized orbitals for solids. It is worthwhile to mention that one of the very first use of the bond polarizabilities as interacting units for the decryption of London dispersion forces has been suggested as early as in 1969 by Claverie and Rein \cite{Claverie:69}; see also \cite{Claverie:78}. 

 Our approach, at least in its simplest form, is situated somewhere between classical models and fully quantum local correlation methods and can be considered (practically in all its forms) as a coarse-grained nonlocal dispersion functional formulated exclusively on the basis of ground state densities and occupied orbitals.
It will be shown how the various matrix elements can be expressed from occupied  orbital quantities only. As a numerical illustration, molecular C$_6$ dispersion coefficients will be calculated from localized orbital contributions and compared to experimental reference data. The paper will be closed by a discussion of  possible future developments.

\section{Theory}

\subsection{Projected oscillator orbitals}

The \textit{a posteriori} localization of the subspace of the occupied orbitals is a relatively standard procedure, which can be achie\-ved following a large variety of localization criteria (for a succinct overview, see Ref. \cite{Alcoba:06}). In the context of correlation energy calculations, \textit{i.e.}\ in various ``local correlation approaches'', the most widely used localization methods are based either on the criterion of Foster and Boys \cite{Foster:60a} or that proposed by  Pipek and Mezey \cite{Pipek:89}. For reasons which become clearer below, in the present work we will use the Foster-Boys localization criterion, which can be expressed in various equivalent forms \cite{Boys:66}. In its the most suggestive formulation, the Foster-Boys' localization procedure consists in the maximization of the squared distance between the centroids of the orbitals:

\begin{align}
 \text{max}\left\{
  \sum_{i<j}^\text{occ}\, \vert
  \bra{\phi_i}\hat{\b{r}}\ket{\phi_i}
- \bra{\phi_j}\hat{\b{r}}\ket{\phi_j}\vert^2
 \right\}
.\end{align}
Another form of the localization criterion, which is strictly equivalent to the previous one, corresponds to the minimization of the sum of quadratic orbital spreads

\begin{align}
 \text{min}\left\{
  \sum_i^\text{occ}\,
       \bra{\phi_i}\hat{\b{r}}^2\ket{\phi_i}
 -\vert\bra{\phi_i}\hat{\b{r}}  \ket{\phi_i}\vert^2
 \right\}
.\end{align}
As it has been demonstrated by Resta \cite{Resta:06b}, the previous minimization 
implies that the sum of the spherically averaged squared off-diagonal matrix elements of the position operator is minimal too. This last property of the Boys' localized orbitals is going to be useful in the development of the present model.

Any set of localized orbitals obtained by a unitary transformation from a set of occupied orbitals
spans the same invariant subspace as the generalized Kohn-Sham
operator, $\hat{f}^\mu$,
and satisfies the equation:

\begin{align}
  \hat{f}^\mu\phi_i^\mu = \sum_j^\text{occ}\,\varepsilon_{ij}^\mu\phi_j^\mu,
\end{align}
where $\hat{f}^\mu\phi_i = \hat{t}+\hat{v}_\text{ne} + \hat{v}_\text{H}
+\hat{v}_{x,\text{HF}}^{\text{lr},\mu} + \hat{v}_{xc,\text{DFA}}^{\text{sr},\mu}$ is 
the range-separated hybrid operator. In this expression $\hat{t}$ is the kinetic energy, $\hat{v}_\text{ne}$ is the nuclear attraction, 
$ \hat{v}_\text{H}$ is the full-range Hartree, $\hat{v}_{x,\text{HF}}^{\text{lr},\mu}$ is the long-range Hartree-Fock (non-local) exchange, and  $\hat{v}_{xc,\text{DFA}}^{\text{sr},\mu}$ is the short-range exchange-correlation potential operator. The range-separation parameter, $\mu$, 
defined below, cuts the electron repulsion terms to  short- and long-
range components. For $\mu=0$ one recovers the full-range density 
functional approximation (DFA), while for $\mu \rightarrow \infty$ one 
obtains the full-range Hartree-Fock theory.

As outlined in the introduction, inspired by the original idea suggested first by Foster and Boys \cite{Foster:60a} and refined later by Boys \cite{Boys:66}, we propose here to construct localized virtual orbitals by multiplying the localized occupied orbital $\phi_i(\b{r})$ by solid spherical harmonics having their origin at the barycenter of the localized occupied orbital. According to  Boys \cite{Boys:66}, this definition can be made independent of the orientation of the coordinate system by choosing local coordinate axes which are parallel to the principal axes of the tensor of the moment of inertia of the charge distribution
$\phi_i^\ast(\b{r}) \phi_i(\b{r})$ (see Appendix~\ref{app:POOlocframe}).

In the following we are going to elaborate the theory for the simplest case, when these \textit{oscillator orbitals} are generated by first order solid spherical harmonic polynomials, \textit{i.e.}\ the $i$-th LMO is multiplied by $(\hat{r}_\alpha - D_\alpha^i)$, where $\hat{r}_\alpha$ is the $\alpha = x,y,z$ component of the position operator and $D^i_\alpha$ is a component of the position vector pointing to the centroid of the $i$-th LMO, defined as $D^i_\alpha=\bra{\phi_i}\hat{r}_\alpha\ket{\phi_i}$.
It is possible to generate oscillator orbitals by higher order spherical harmonics too, which is left for forthcoming work.
 We denote the POO by $\ket{\tilde{\phi}_{i_\alpha}}$, where the index $i_\alpha$ refers to the fact that the OO has been generated from the $i$-th LMO by using the $\hat{r}_\alpha$ function. For the sake of the simplicity of the formulae, the POOs will be expressed in the laboratory frame; the expressions for the dipolar POOs in a local frame are shown in the Appendix~\ref{app:POOlocframe}. The POO reads in the laboratory frame as

\begin{align}
\label{eq:defPOO}
  \ket{\tilde{\phi}_{i_\alpha}}
&= \biggl(\hat{I}-\sum_m^\text{occ}\, \ket{\phi_m}\bra{\phi_m}\biggr)
   \bigl(\hat{r}_\alpha - D_\alpha^i\bigr)\ket{\phi_i}
 =\hat{Q}\,\hat{r}_\alpha \ket{\phi_i}
,\end{align}
with $\hat{Q}=(\hat{I}-\sum_m^\text{occ}\, \ket{\phi_m}\bra{\phi_m})$, the projector onto the virtual space.

\begin{figure*}
\begin{center}
\subfiguretopcaptrue
     \subfigure[O lone pair $\ket{i}$]      {\includegraphics[trim=50mm 80mm 00mm 80mm,clip=true,keepaspectratio=true,width=0.24\textwidth]{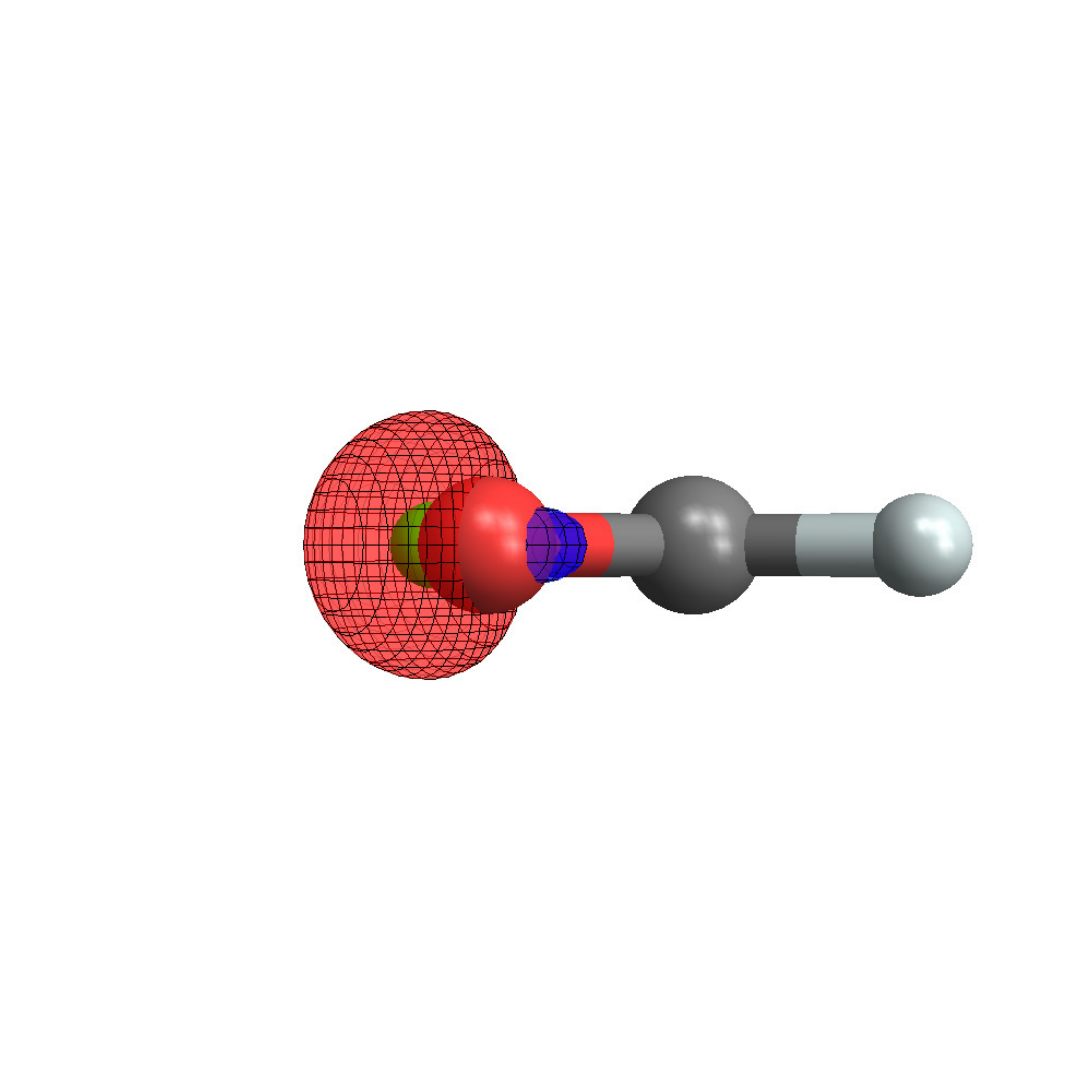}}
     \subfigure[$\hat{Q}\,\hat{r}_x\ket{i}$]{\includegraphics[trim=50mm 80mm 00mm 80mm,clip=true,keepaspectratio=true,width=0.24\textwidth]{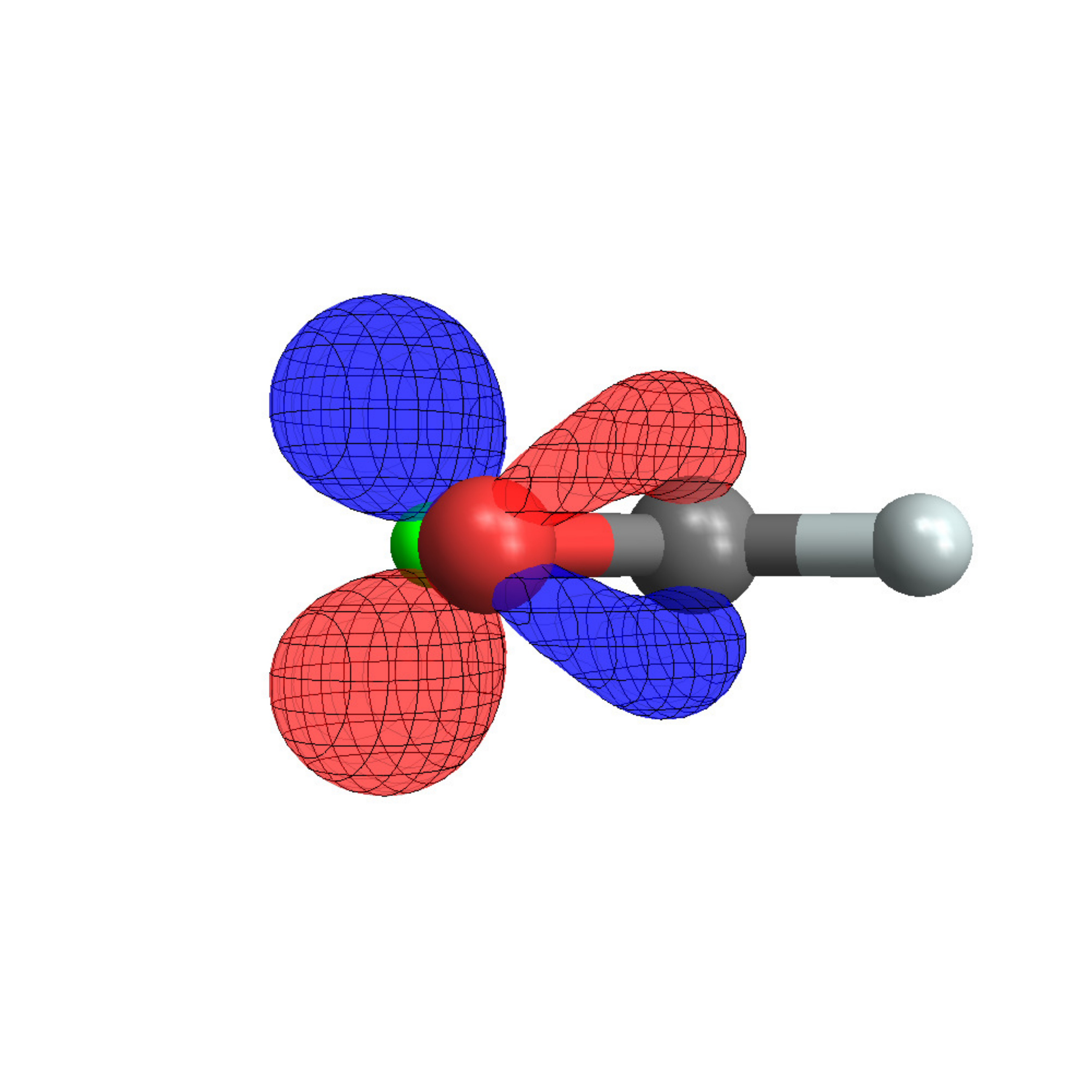}}
     \subfigure[$\hat{Q}\,\hat{r}_y\ket{i}$]{\includegraphics[trim=50mm 80mm 00mm 80mm,clip=true,keepaspectratio=true,width=0.24\textwidth]{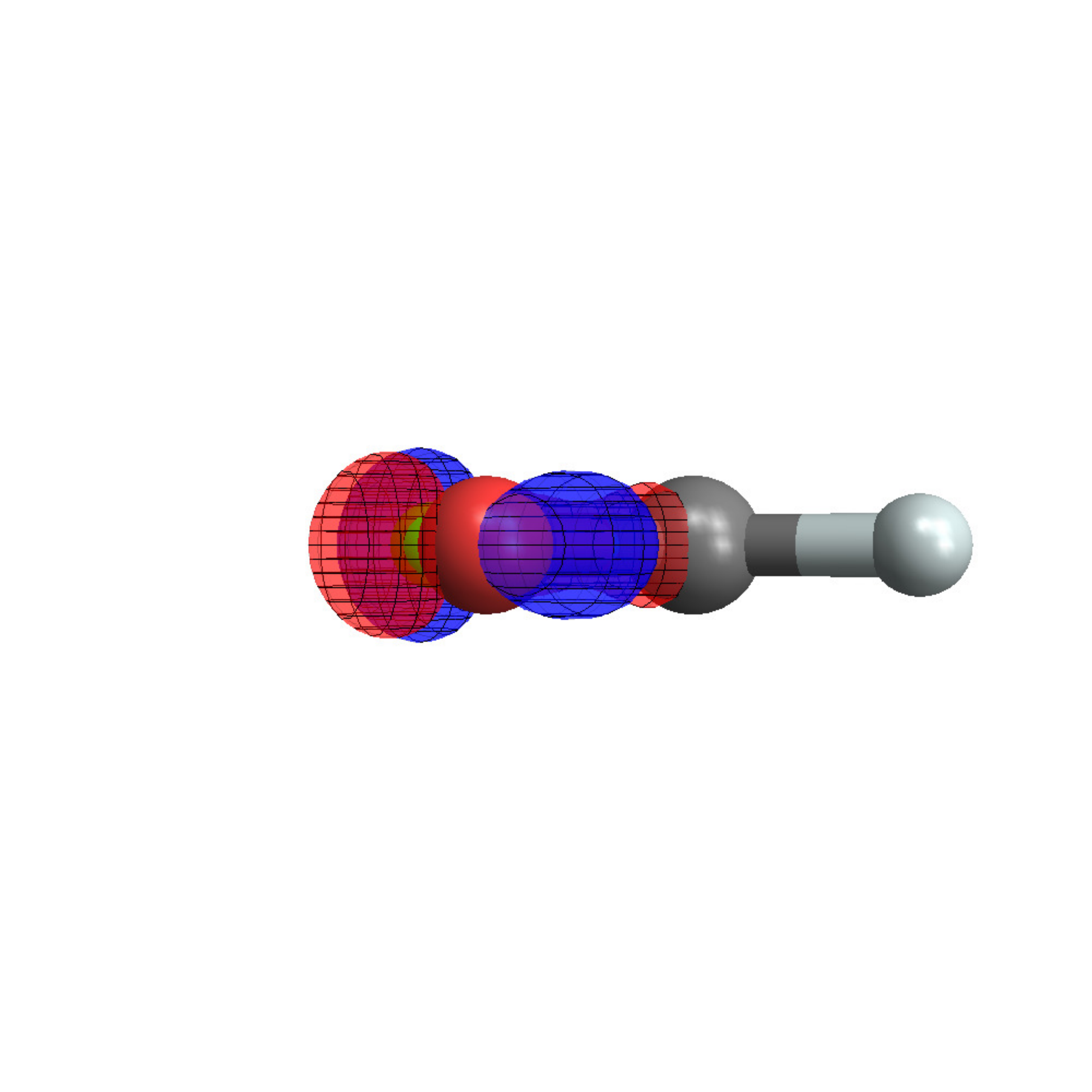}}
     \subfigure[$\hat{Q}\,\hat{r}_z\ket{i}$]{\includegraphics[trim=50mm 80mm 00mm 80mm,clip=true,keepaspectratio=true,width=0.24\textwidth]{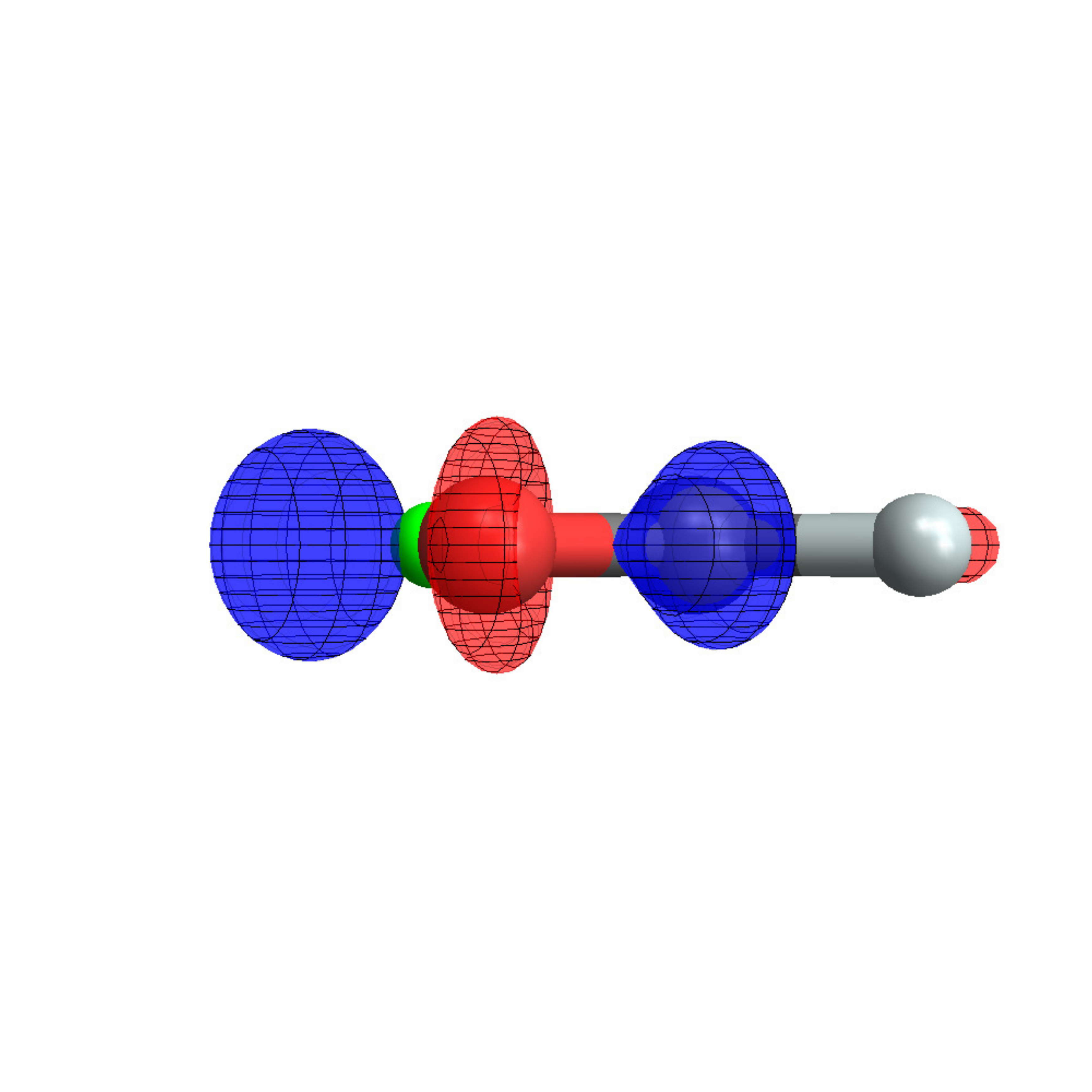}}

     \subfigure{\includegraphics[trim=00mm 00mm 00mm 90mm,clip=true,keepaspectratio=true,width=0.24\textwidth]{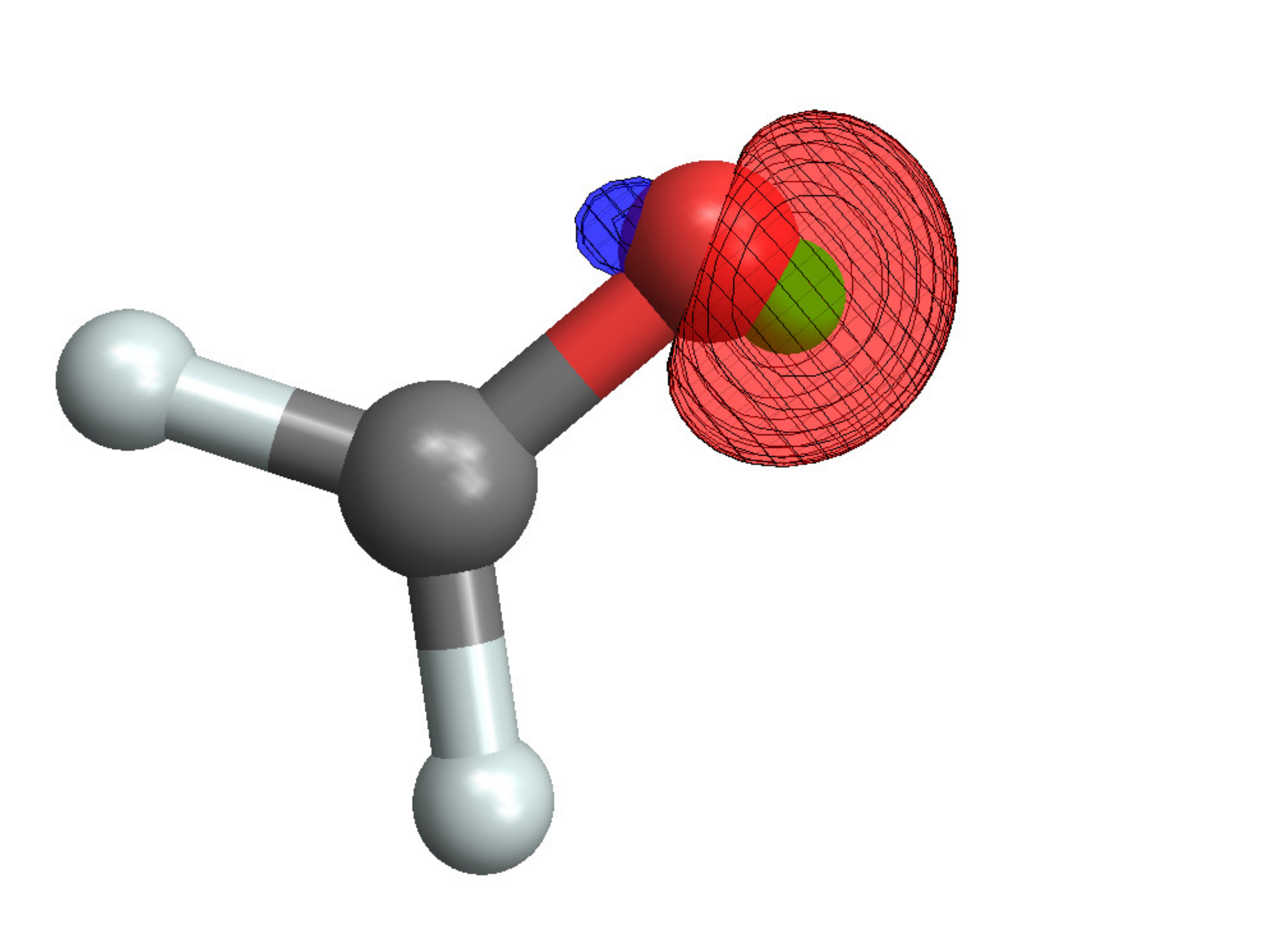}}
     \subfigure{\includegraphics[trim=00mm 00mm 00mm 90mm,clip=true,keepaspectratio=true,width=0.24\textwidth]{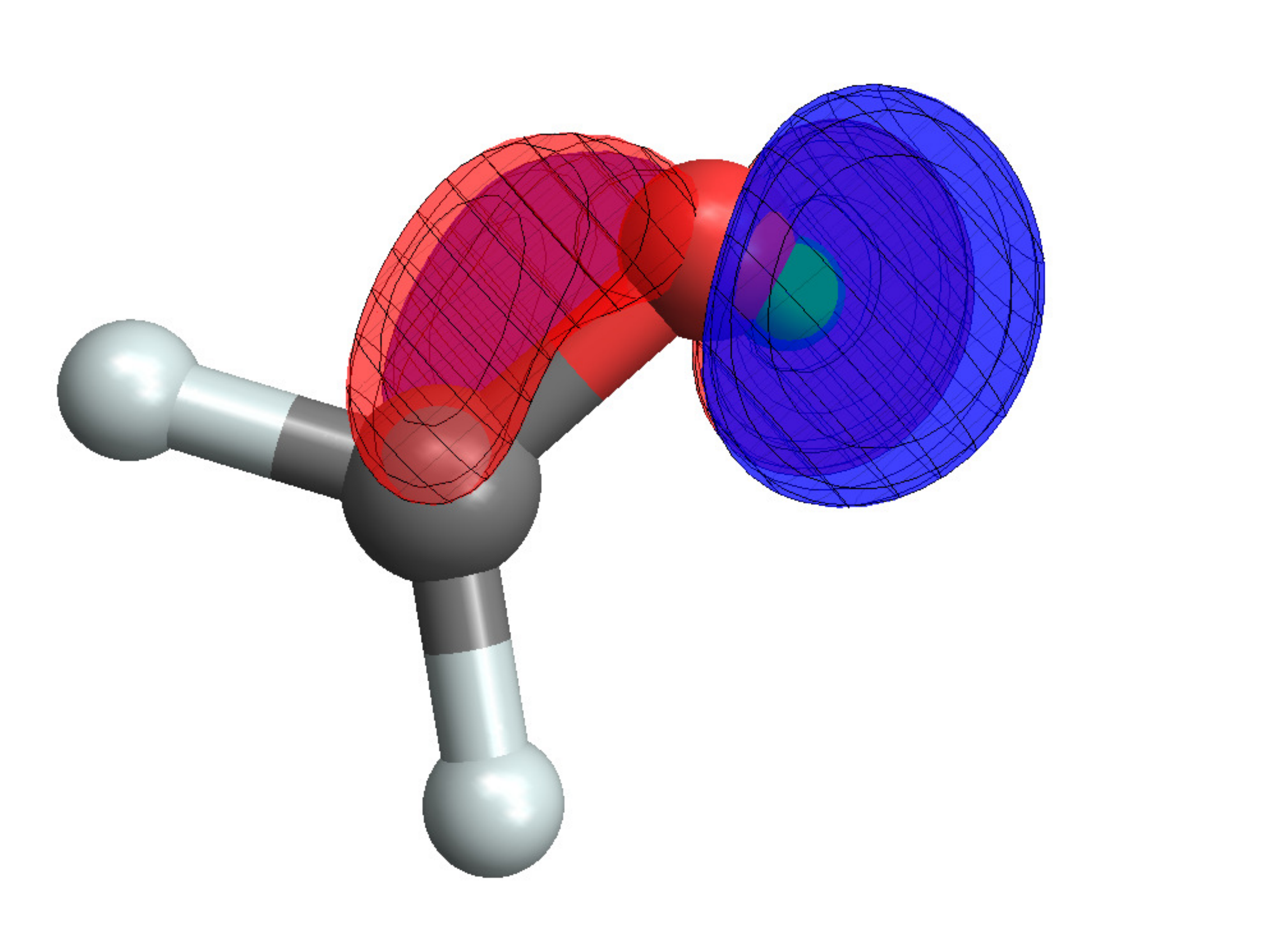}}
     \subfigure{\includegraphics[trim=00mm 00mm 00mm 90mm,clip=true,keepaspectratio=true,width=0.24\textwidth]{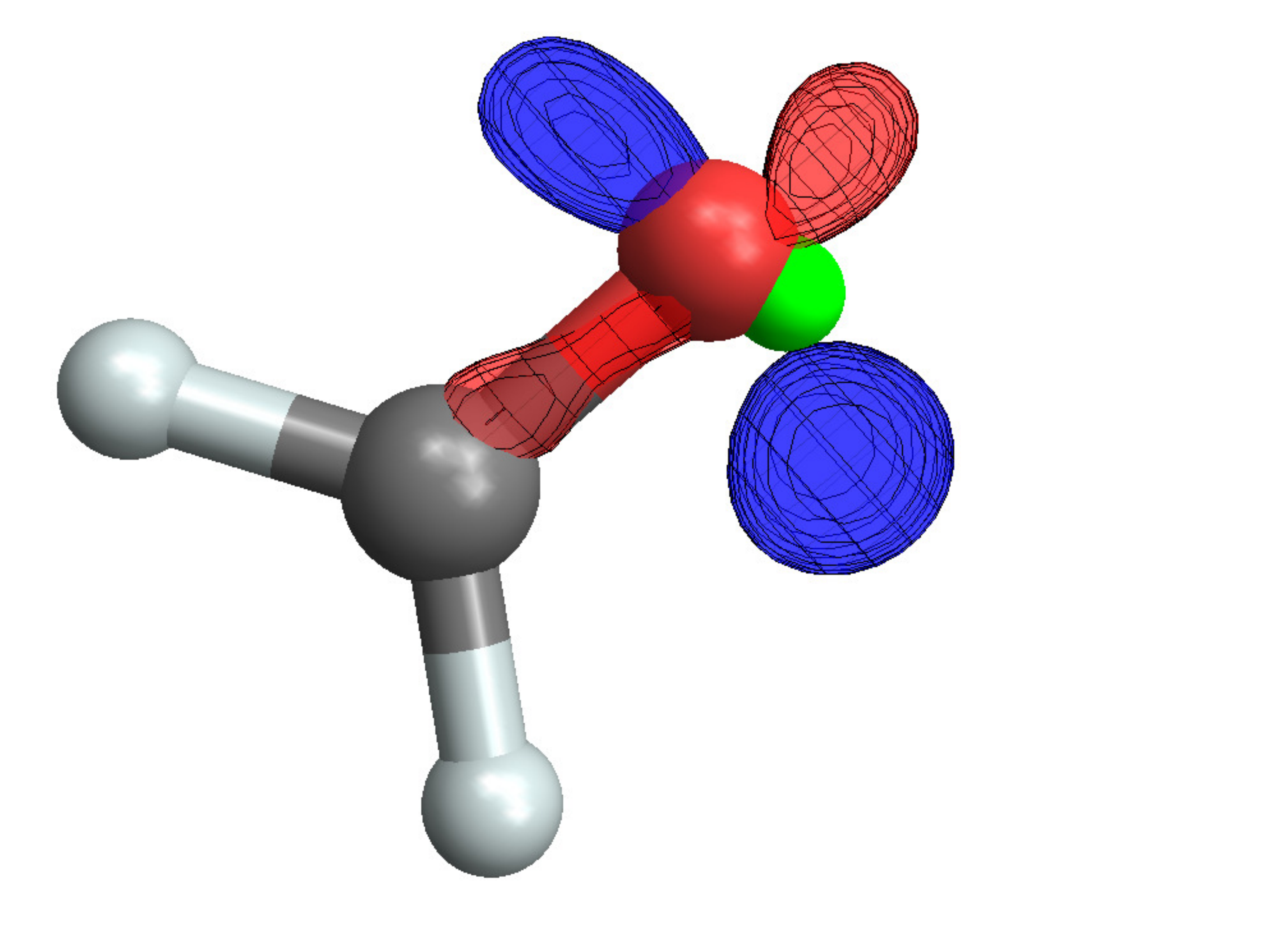}}
     \subfigure{\includegraphics[trim=00mm 00mm 00mm 90mm,clip=true,keepaspectratio=true,width=0.24\textwidth]{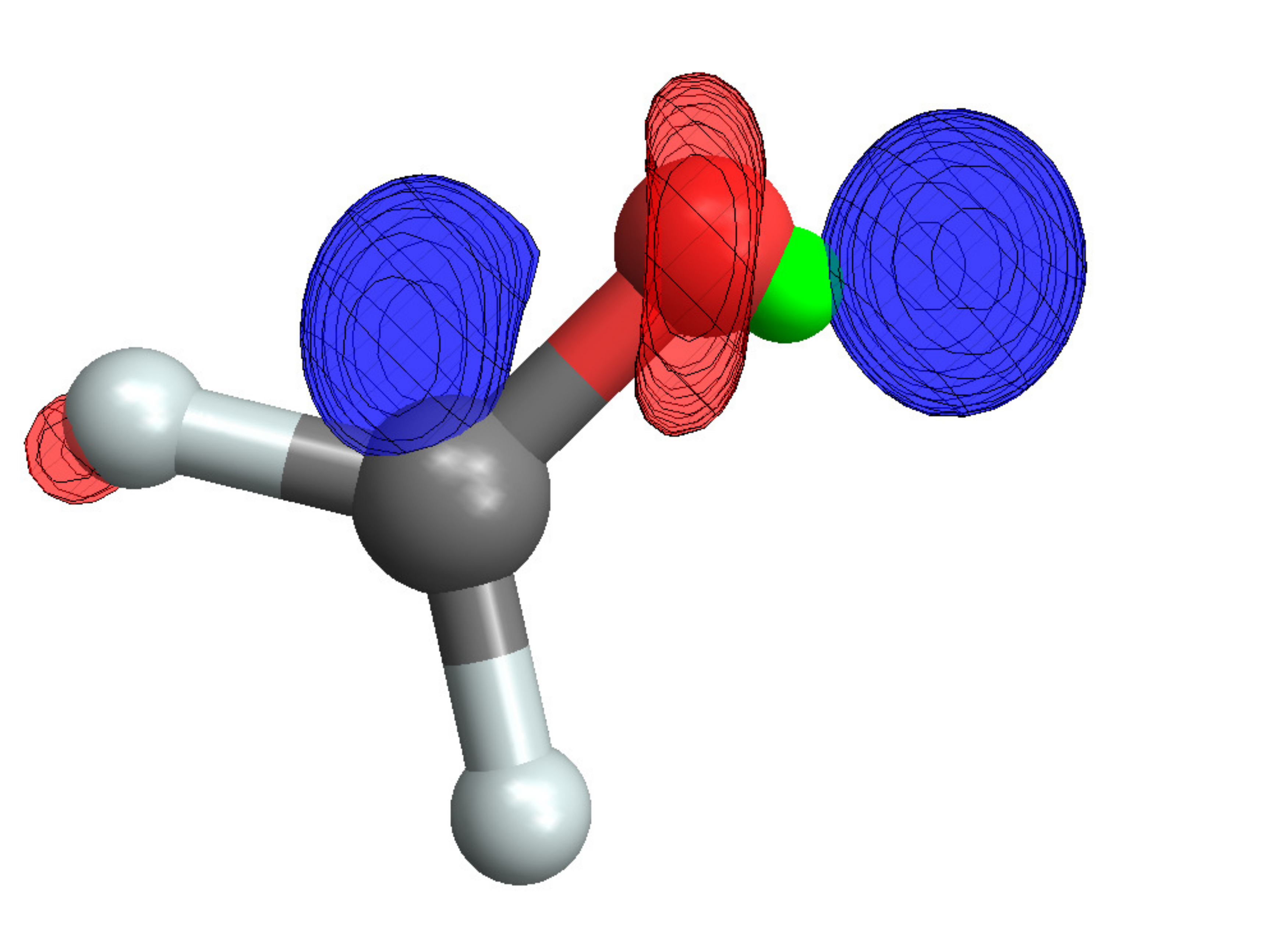}}
\caption{Projected Oscillator Orbitals ((b),(c),(d)) generated by projection of the products of first order solid spherical harmonics polynomials with the O lone pair orbital of \ce{H2C=O} seen in (a).
The harmonics are aligned with the local frame axes, \textit{i.e.}\ the principal axes of the tensor of the moment of inertia of the charge distribution of the oxygen lone pair orbital.
The \textit{green dot} indicates the position of the LMO centroid.
}
\end{center}
\end{figure*}

\begin{figure*}
\begin{center}
\subfiguretopcaptrue
     \subfigure[\ce{C-H} bonding orbital $\ket{i}$]{\includegraphics[trim=30mm 80mm 00mm 80mm,clip=true,keepaspectratio=true,width=0.24\textwidth]{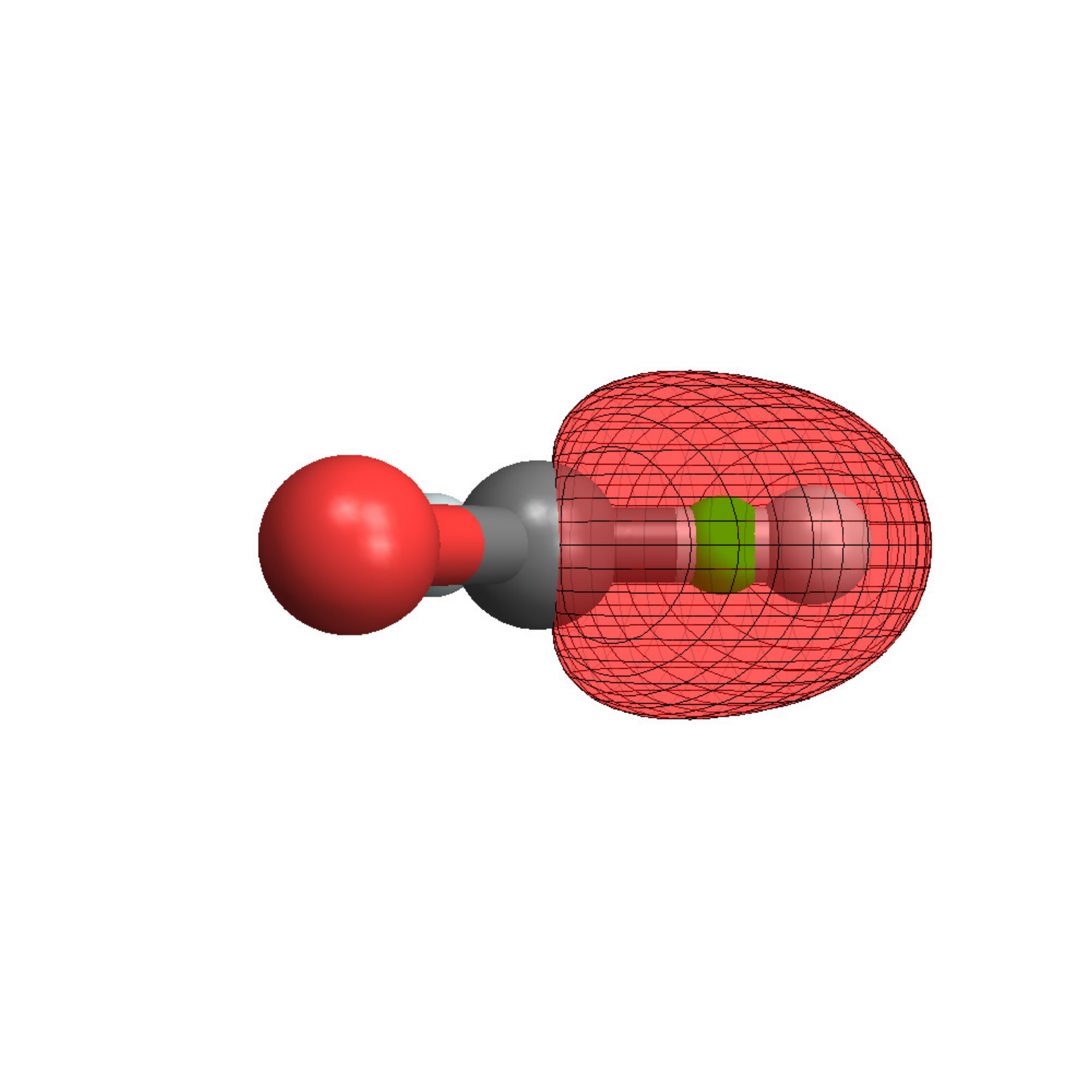} }
     \subfigure[$\hat{Q}\,\hat{r}_x\ket{i}$]       {\includegraphics[trim=30mm 80mm 00mm 80mm,clip=true,keepaspectratio=true,width=0.24\textwidth]{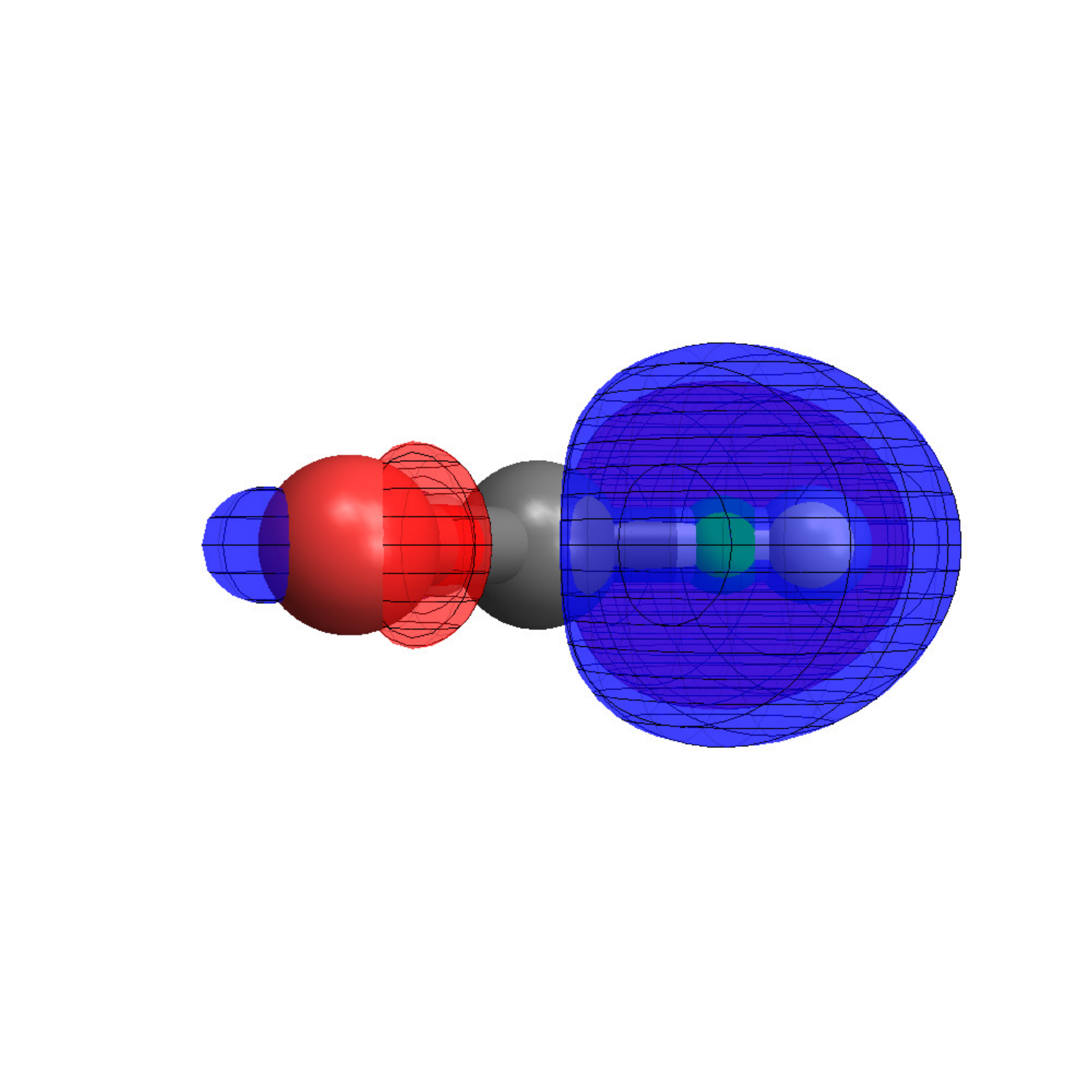}}
     \subfigure[$\hat{Q}\,\hat{r}_y\ket{i}$]       {\includegraphics[trim=30mm 80mm 00mm 80mm,clip=true,keepaspectratio=true,width=0.24\textwidth]{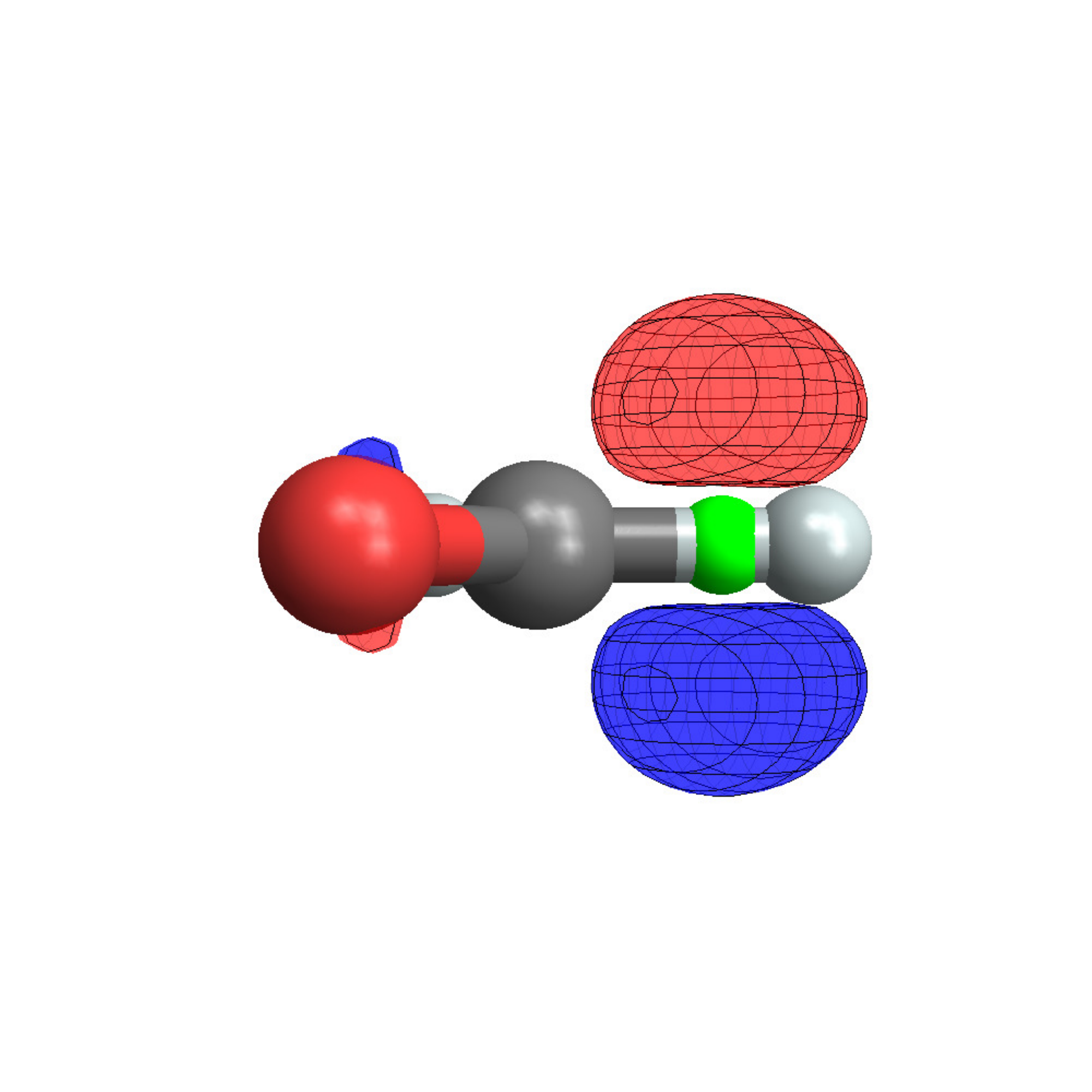}}
     \subfigure[$\hat{Q}\,\hat{r}_z\ket{i}$]       {\includegraphics[trim=30mm 80mm 00mm 80mm,clip=true,keepaspectratio=true,width=0.24\textwidth]{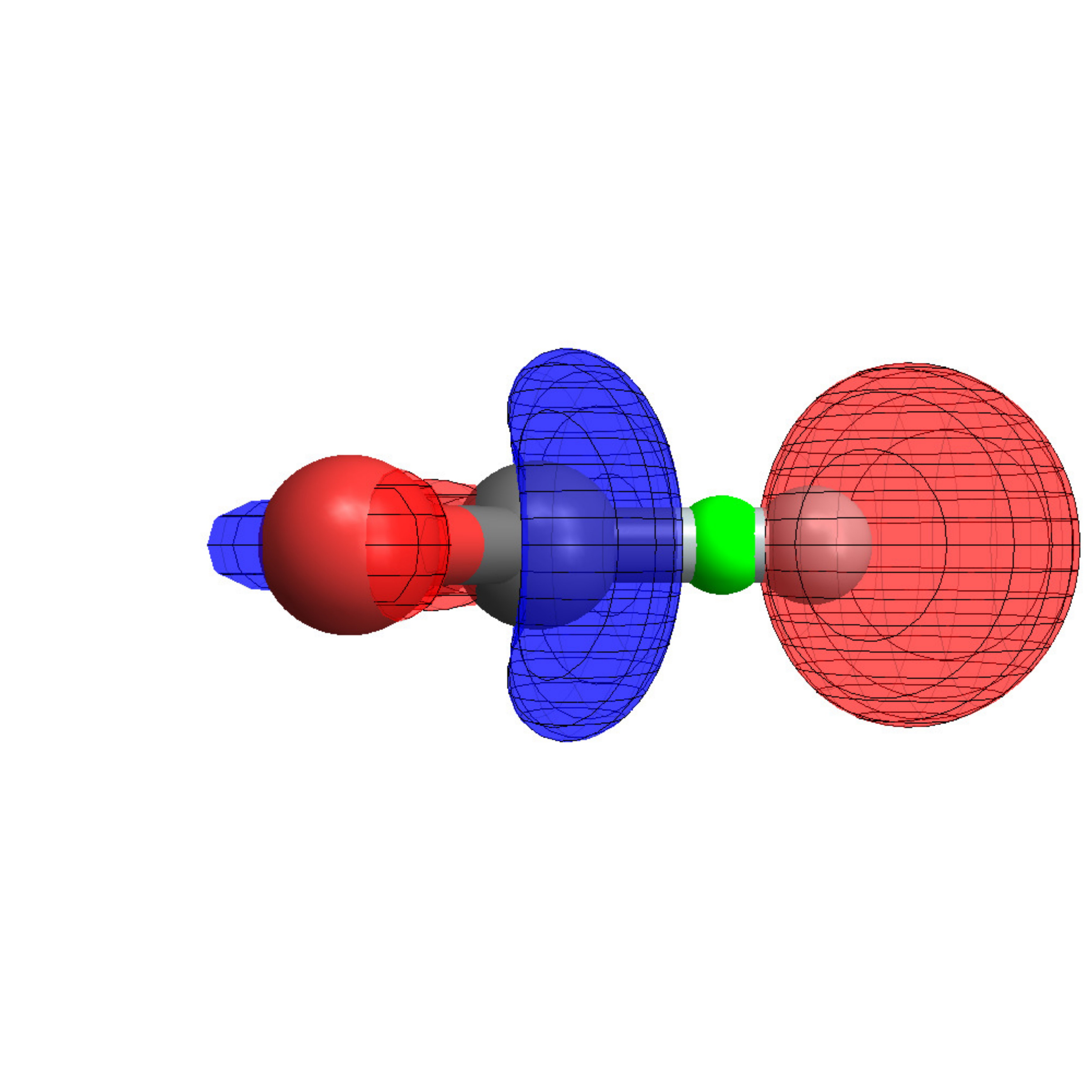}}

     \subfigure{\includegraphics[trim=50mm 40mm 00mm 40mm,clip=true,keepaspectratio=true,width=0.24\textwidth]{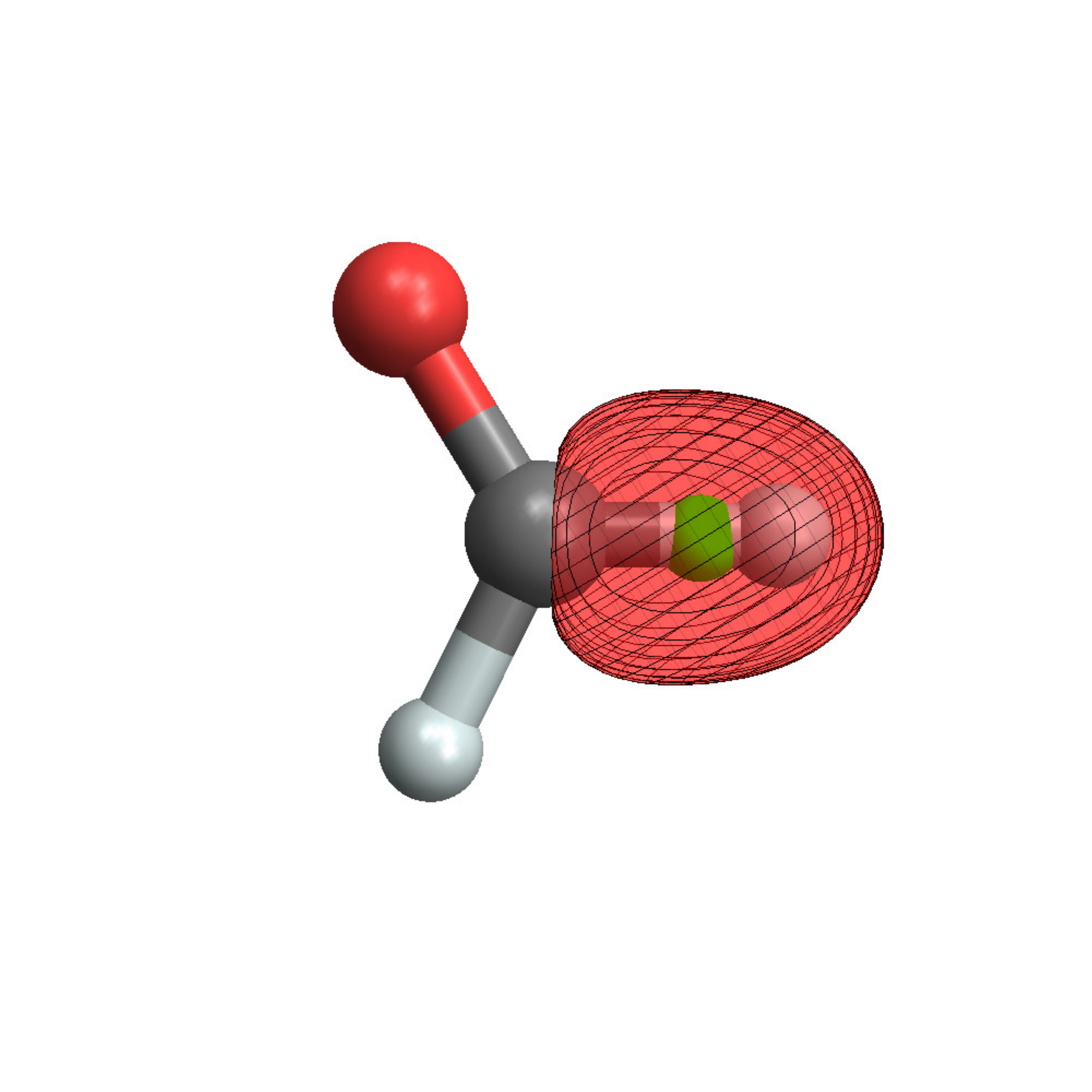} }
     \subfigure{\includegraphics[trim=50mm 40mm 00mm 40mm,clip=true,keepaspectratio=true,width=0.24\textwidth]{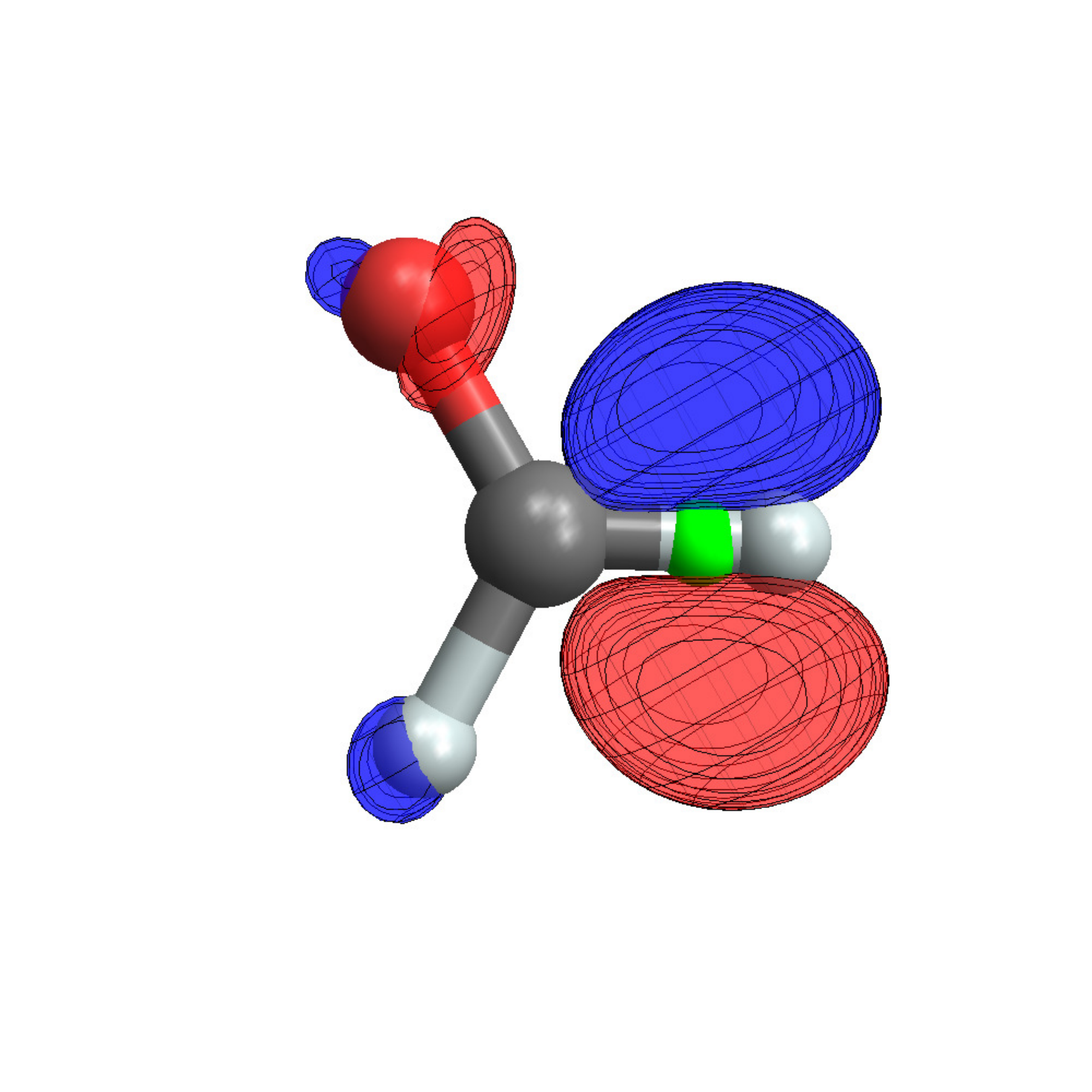}}
     \subfigure{\includegraphics[trim=50mm 40mm 00mm 40mm,clip=true,keepaspectratio=true,width=0.24\textwidth]{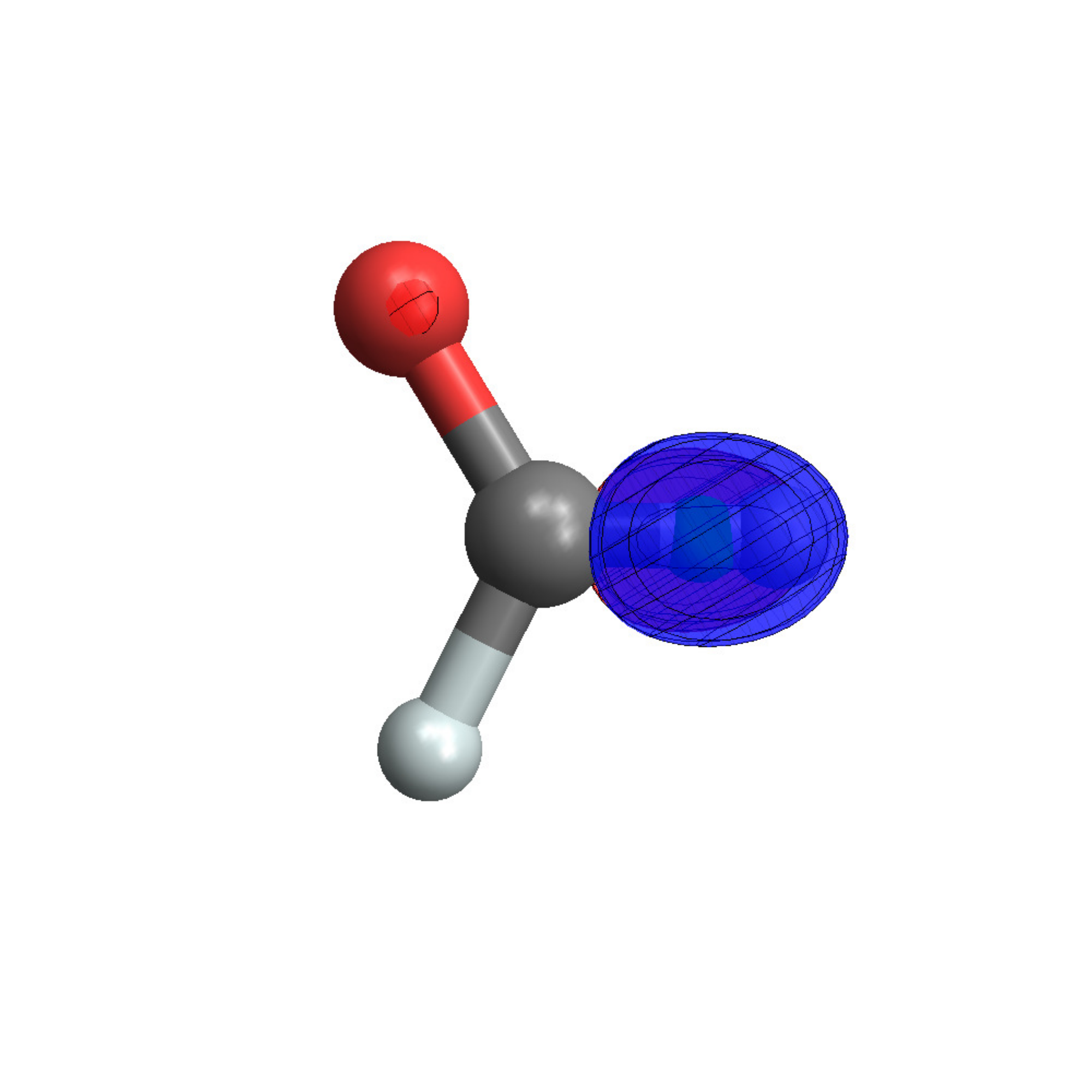}}
     \subfigure{\includegraphics[trim=50mm 40mm 00mm 40mm,clip=true,keepaspectratio=true,width=0.24\textwidth]{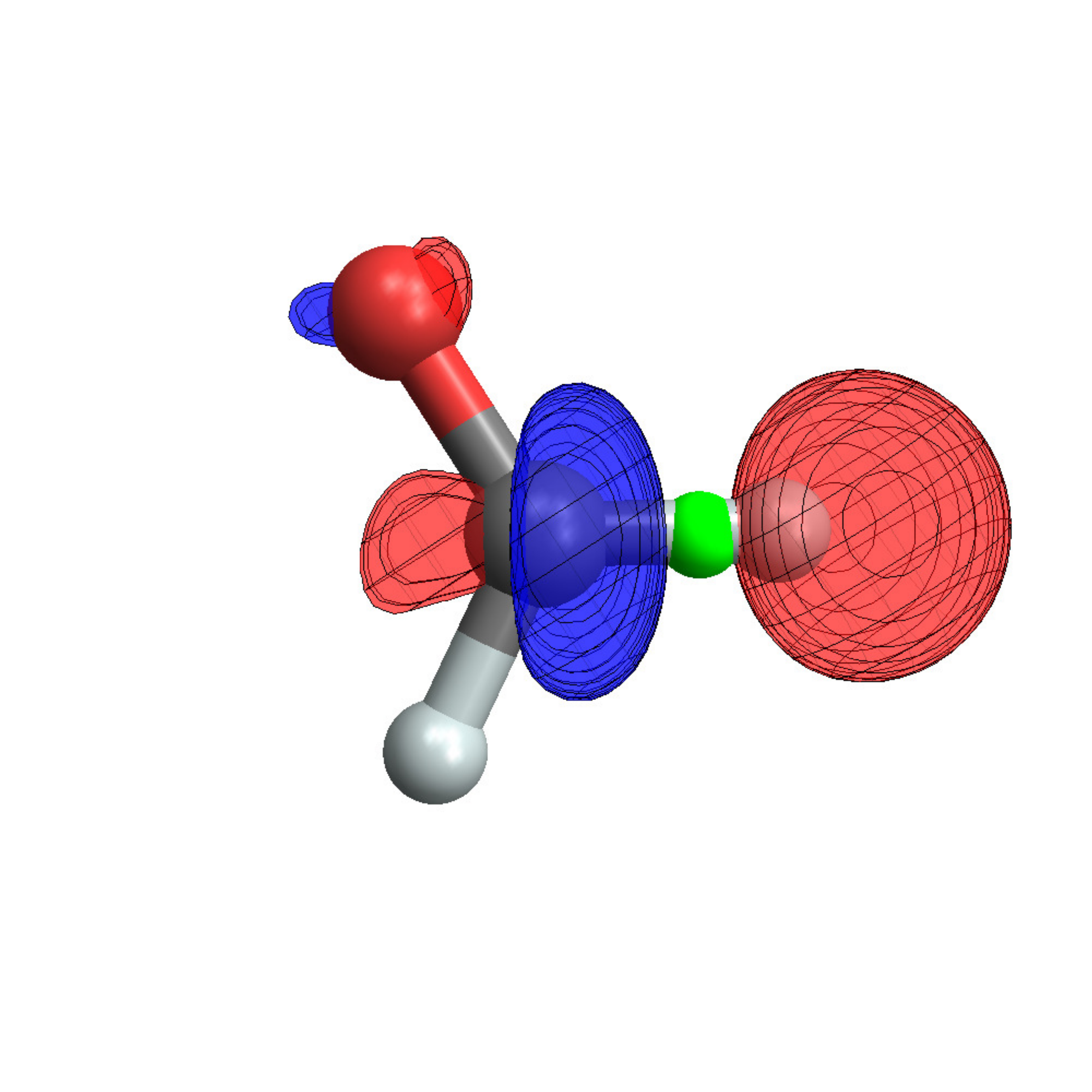}}
\caption{Projected Oscillator Orbitals ((b),(c),(d)) generated by projection of the products of first order solid spherical harmonics polynomials with the \ce{C-H} bonding orbital of \ce{H2C=O} (a).
The harmonics are aligned with with the local frame axes, \textit{i.e.}\  the principal axes of the tensor of the moment of inertia of the charge distribution of the \ce{C-H} bonding orbital.
The \textit{green dot} indicates the position of the LMO centroid.
}
\end{center}
\end{figure*}

In Eq.~\ref{eq:defPOO}  the ``pure'' (OO) component of the POO is
$\hat{r}_\alpha\ket{\phi_i}$,
while the orthogonalization tails stem from the term
$-\sum_{m\neq i}^\text{occ}\, \ket{\phi_m}\bra{\phi_m}\hat{r}_\alpha\ket{\phi_i}$.
In this sense the  ``locality'' of the OO is somewhat deteriorated, since we have contributions from each of the other LMOs. At this point  the Boys localization criterion will be at our advantage, since it ensures that the sum of the off-diagonal elements of the $x, y$ and $z$ operators taken between the occupied orbitals, and appearing in the orthogonalization tails, be minimized \cite{Resta:06b}. In this sense, the Boys-localization scheme seems to be naturally adapted for the construction of dipolar oscillator orbitals.

To illustrate the concept of dipolar oscillator orbitals, two examples are taken from the oxygen lone pair and the C-H bonding orbitals of the formaldehyde molecule, described in the above-defined local frame. Figs 1 and 2 show the localized orbitals and the three projected dipolar oscillator orbitals having a nodal surface intersecting the orbital centroid and oriented in the three Cartesian coordinate directions of the local coordinate system, $x$, $y$ and $z$. It is quite clear that the node coincides with the region of the highest electron density of the orbital and in this sense ensures an optimal description of the correlation. Higher order polynomials generate virtual orbitals with further nodes.

From now on, we will use the simplified notations $\ket{\phi_i}\equiv\ket{i}$ and $\ket{\tilde{\phi}_{i_\alpha}}\equiv\ket{i_\alpha}$, in other words the subscript $\alpha$ on the orbital index indicates that it is an oscillator orbital.
We designate the occupied (canonical or localized) molecular orbitals as $i,j,k,\dots$ and the canonical virtual molecular orbitals (VMOs) as $a,b,c,\dots$.

The oscillator orbitals are non-orthogonal among each other; their overlap integral can be evaluated
using the idempotency of the projectors:

\begin{align}
\label{eq:overlap}
S_{i_\alpha,j_\beta}
& = \braket{i_\alpha}{j_\beta}
 = \bra{i}\hat{r}_\alpha \hat{r}_\beta\ket{j} - \sum_m^\text{occ}\, \bra{i}\hat{r}_\alpha\ket{m}\bra{m}\hat{r}_\beta\ket{j}
.\end{align}
Note that the overlap matrix $\b{S}$ is of size $N_\POO\times N_\POO$, where $N_\POO$ is the number of projected oscillator orbitals.

The POOs can be expanded in terms of a set of orthonormalized virtual orbitals (\textit{e.g.}\ the set of canonical virtuals). Although later we  eliminate  explicit reference to the set of virtual orbitals of the Fock/Kohn-Sham operator (\textit{i.e.}\ everything will be written only in terms of the occupied orbitals or equivalently in terms of the corresponding density matrix), with the help of the resolution of identity, we  give the explicit form of the coefficient matrix linking the POOs with the virtuals:

\begin{align}
\label{eq:virt2poo}
 \ket{i_\alpha}
& = \biggl(\sum_p^\text{all}\, \ket{p}\bra{p} \biggr)\hat{Q}\,\hat{r}_\alpha\ket{i}
  = \sum_a^\text{virt}\, \ket{a}\bra{a}\hat{r}_\alpha\ket{i}
  = \sum_a^\text{virt}\, \ket{a}\, V_{a\,i_\alpha}
.\end{align}
The matrix  $\b{V}$ is constructed simply from the elements of the occupied/virtual block of the position operator, $ V_{a\,i_\alpha}=\bra{a}\hat{r}_\alpha\ket{i}$. The overlap matrix of the expanded POOs can be written in terms of the coefficient matrix $\b{V}$:

\begin{align}
S_{i_\alpha,j_\beta}
& = \braket{i_\alpha}{j_\beta}
  = \sum_{ab}^\text{virt}\, V^\dagger_{i_\alpha\,a}\,\braket{a}{b}\,V^{\vphantom{\dagger}}_{b\,j_\beta}
  = ( \b{V}^\dagger\,\b{V} )_{i_\alpha j_\beta}
.\end{align}

Higher order oscillator orbitals, not used in the present work, can be generated in an analogous manner, using higher order solid spherical harmonic functions.

\subsection{Ring CCD-RPA equations with POOs}

In the ring CCD (ring coupled cluster double excitations) formulation \cite{Toulouse:11}, the general RPA correlation energy (direct-RPA or RPA-exchange) is a sum of pair-contributions attributed to a pair of occupied (localized) orbitals:

\begin{align}
\label{eq:EcRPA}
E_c^\text{RPA} = \frac{1}{2} \sum_{ij}^\text{occ}\, \text{tr}\,\bigl\{\b{B}^{ij}\,\b{T}^{ij}\bigr\}
,\end{align}
where the amplitudes $\b{T}^{ij}$ satisfy the Riccati equations, which can be written in terms of orthogonalized occupied $i,j,k,\ldots$  and virtual $a,b,c,\ldots$ orbitals as:

\begin{align}
\label{eq:RiccatiMO}
 \b{R}^{ij} =  \b{B}^{ij} + ((\pmb{\epsilon}+\b{A})\b{T})^{ij} + (\b{T}(\b{A}+\pmb{\epsilon}))^{ij} + (\b{T}\b{B}\b{T})^{ij}=\b{0}
,\end{align}
with the matrix elements:

\begin{align}
 \label{eq:defDeltaEpsAB}
 \epsilon^{ij}_{ab} &= \delta_{ij} f_{ab} - f_{ij} \delta_{ab}
\nonumber\\
 A^{ij}_{ab} &=K^{ij}_{ab}-J^{ij}_{ab}=\braket{aj}{ib}\!-\!\braket{aj}{bi}
\\
 B_{ab}^{ij} &=K^{ij}_{ab}-K^\prime{}^{ij}_{ab}=\braket{ij}{ab}\!-\!\braket{ij}{ba}
\nonumber
,\end{align}
where $\b{f}$ is the fock matrix of the fock operator $\hat{f}$ and with the two-electron integrals of spin-or\-bitals written with the physicists' notation:

\begin{align}
\label{eq:2elint}
\braket{ij}{ab}=\int \phi_i^\ast(\b{r})\phi_j^\ast(\b{r}^\prime)w(\b{r},\b{r}^\prime)\phi_a(\b{r})\phi_b(\b{r}^\prime) \; d\b{r}d\b{r}^\prime
,\end{align}
where $w(\b{r},\b{r}^\prime)=\vert\b{r}^\prime-\b{r}\vert^{-1}$ is the Coulomb electron repulsion interaction.

Note that the size of the matrices in Eq.~\ref{eq:EcRPA} and~\ref{eq:RiccatiMO} is, for each pair $[ij]$, $N_\text{virt}\times N_\text{virt}$
and that in this notation the matrix multiplications are understood as, for example:  $(\b{T}(\b{A}+\pmb{\epsilon}))^{ij}_{ab}=\sum_{mc}\, T^{im}_{ac} A^{mj}_{cb}+T^{im}_{ac} \epsilon^{mj}_{cb}$.

Using the transformation rule between the amplitudes in the VMOs and in the POO basis, $\b{T}^{ij}=\b{V}\b{T}^{ij}_\text{POO}\b{V}^\dagger$ (see Appendix~\ref{app:ricPOO}),
the Riccati equations can be recast in the POO basis as (again, see Appendix~\ref{app:ricPOO}):

\begin{align}
\label{eq:RiccatiPOO}
\b{R}^{ij}_\POO
&= \b{B}^{ij}_\POO
 + (\pmb{\epsilon}_\POO+\b{A}^{im}_\POO)\,\b{T}_\POO^{mj}\,\b{S}^{~}_\POO
\nonumber\\&\quad
 + \b{S}^{~}_\POO\,\b{T}_\POO^{im}\,(\pmb{\epsilon}_\POO+\b{A}^{mj}_\POO)
\nonumber\\&\quad
 + \b{S}^{~}_\POO\,\b{T}_\POO^{im}\,\b{B}^{mn}_\POO\,\b{T}_\POO^{nj}\,\b{S}^{~}_\POO=\b{0}
.\end{align}
This corresponds to a local formulation of the ring CCD amplitudes equations, and the dimension of the matrices, emphasized by the subscript POO, is merely $N_\POO\times N_\POO$.
As explained in Appendix~\ref{app:WorkRic}, this type of Riccati equations can be solved iteratively in a pseudo-canonical basis.

\subsection{Local excitation approximation}

Since the excitations are limited to ``pair-domains'', the effective dimension of the equations for a pair is actually rou\-ghly independent from the size of the system, just like in any local correlation procedure.

As the simplest approximation, one can take only the three excitations to the dipolar POOs generated by a selected LMO, \textit{i.e.}\ for each pair  of LMOs  $[ij]$ we have the local excitations  $i\rightarrow  i_\alpha$ and  $j\rightarrow  j_\beta$, leading to a $3\times 3$ problem to solve and iterate on.
Within this approximation, all matrices involved in the derivations can be fully characterized by two occupied LMO indices and two cartesian components, $\alpha$ and $\beta$ (the subscript POO is omitted from now on):

\begin{align}
(\b{A})^{ij}_{i_\alpha j_\beta}&\equiv(\b{A})^{ij}_{\alpha\beta}&\text{and: }&&(\b{S})_{i_\alpha j_\beta}     &\equiv(\b{S})^{ij}_{\alpha\beta}
\nonumber\\
(\b{B})^{ij}_{i_\alpha j_\beta}&\equiv(\b{B})^{ij}_{\alpha\beta}&            &&(\b{f})_{i_\alpha j_\beta}     &\equiv(\b{f})^{ij}_{\alpha\beta}
\\
(\b{T})^{ij}_{i_\alpha j_\beta}&\equiv(\b{T})^{ij}_{\alpha\beta}
\nonumber
\end{align}

With this in mind, and with the additional approximation which consists in neglecting the overlap between POOs coming from different LMOs, i.e.\
$(\b{S})^{ij}_{\alpha\beta} \approx \delta_{ij}\,(\b{S})^{ii}_{\alpha\beta}$, the direct RPA Riccati equations of Eq.~\ref{eq:RiccatiPOO} become:

\begin{align}
\label{eq:RiccatiLocal}
  \b{R}^{ij}
&=\b{B}^{ij}
 +\b{f}^{ii}\,\b{T}^{ij}\,\b{S}^{jj}
 -   f _{ii}\,\b{S}^{ii}\,\b{T}^{ij}\,\b{S}^{jj}
 +\b{A}^{im}\,\b{T}^{mj}\,\b{S}^{jj}
\nonumber\\&\qquad\;\,
 +\b{S}^{ii}\,\b{T}^{ij}\,\b{f}^{jj}
 -\b{S}^{ii}\,\b{T}^{ij}\,\b{S}^{jj}\,f _{jj}
 +\b{S}^{ii}\,\b{T}^{im}\,\b{A}^{mj}
\nonumber\\&\qquad\;\,
 +\b{S}^{ii}\,\b{T}^{im}\,\b{B}^{mn}\,\b{T}^{nj}\,\b{S}^{jj}
=\b{0}
.\end{align}
In the above equation, we have written explicitly the fock matrix contributions and used implicit summation conventions over $m$ and $n$.
A detailed derivation of Eq.~\ref{eq:RiccatiLocal} from Eq.~\ref{eq:RiccatiPOO} is shown in Appendix~\ref{app:RiccLoc}.
These Riccati equations can be solved by a transformation to the pseudo-canonical basis, as described in Appendix~\ref{app:WorkRic}.

\subsection{Multipole approximation for the long-range two-electron integrals in the POO basis}

In the context of range-separation, and in the spirit of constructing an \textit{approximate} theory which takes advantage of the localized character of the occupied molecular orbitals,
we are going to proceed via a multipole expansion of the long-range two-electron integrals.

The  matrices $\b{A}^{ij}_\POO$ and $\b{B}^{ij}_\POO$ will be reinterpreted  
in terms of long-range two-electron integrals, \textit{i.e.}\ $w(\b{r},\b{r}^\prime)$
will be replaced by 
$w_\text{lr}(\b{r},\b{r}^\prime)= 
\text{erf}\bigr(\mu\vert\b{r}^\prime-\b{r}\vert\bigl)\vert\b{r}^\prime-\b{r}\vert^{-1}$
in Eq.~\ref{eq:2elint}.
They read respectively as (see Eq.~\ref{eq:defDeltaEpsAB} and the transformation described in Appendix~\ref{app:ricPOO}):

\begin{align}
&K^{ij}_{m_\alpha n_\beta}-J         ^{ij}_{m_\alpha n_\beta} = 
\braket{m_\alpha j}{i n_\beta}_\text{lr}-
\braket{m_\alpha j}{n_\beta i}_\text{lr}
\\
&K^{ij}_{m_\alpha n_\beta}-K^\prime{}^{ij}_{m_\alpha n_\beta} = 
\braket{ij}{m_\alpha n_\beta}_\text{lr}-
\braket{ij}{n_\beta m_\alpha}_\text{lr}
.\end{align}

Note that these integrals could be calculated by using the POO to VMO transformation of Eq.~\ref{eq:virt2poo}. However, 
such an expression is not in harmony with our goal of getting rid of virtual orbitals, since it requires the full set of integrals transformed in occupied and canonical VMOs with an additional two-index transformation. We could formally eliminate virtual molecular orbitals by applying the resolution of identity, but in this case we would be faced with new type of two-electron integrals, in addition to the usual ones generated by the Coulomb interaction $\vert\b{r}-\b{r}^\prime\vert^{-1}$, namely integrals generated by $\hat{r}_\alpha\,\vert\b{r}-\b{r}^\prime\vert^{-1}$, $\hat{r}_\beta^\prime\,\vert\b{r}-\b{r}^\prime\vert^{-1}$ and
$\hat{r}_\alpha\,\hat{r}_\beta^\prime\vert\b{r}-\b{r}^\prime\vert^{-1}$. 
Therefore we are going to proceed by a multipole expansion technique.

The expansion centre for the multipole expansion will be chosen at the centroid of the LMOs, \textit{i.e.}\ in this example at $\b{D}^i$ and $\b{D}^j$. 
Using the second order long-range interaction tensor $\b{L}^{ij}(\b{D}^{ij})$, with $\b{D}^{ij}=\b{D}^i-\b{D}^j$ (see Appendix~\ref{app:LRT}), we have

\begin{align}
K^{ij}_{m_\alpha n_\beta}=
   \braket{m_\alpha j}{i n_\beta}_\text{lr}
&\approx\sum_{\gamma\delta}\,
   \bra{m_\alpha}\hat{r}_\gamma\ket{i}\,
   L^{ij}_{\gamma\delta}\,
   \bra{j}\hat{r}_\delta\ket{n_\beta}
\nonumber\\&\quad
 + \text{higher multipole terms}
.\end{align}

A truly remarkable formal result emerging from the framework of oscillator orbitals  is that
the $\hat{r}_\gamma$ matrix element between the POO $ m_\alpha$ and the LMO $i$ that appears in the previous equation
is nothing else but the overlap between the POOs  $ m_\alpha$ and $ i_\gamma$ (cf.~Eq.~\ref{eq:overlap}):

\begin{align}
\label{eq:remarkable}
\bra{m_\alpha}\hat{r}_\gamma\ket{i}
&=\bra{m}\hat{r}_\alpha\,\hat{r}_\gamma\ket{i}-
  \sum_n^\text{occ}\,
  \bra{m}\hat{r}_\alpha\ket{n}\bra{n}\hat{r}_\gamma\ket{i}
= S_{m_\alpha i_\gamma}
,\end{align}
so that the matrix element $K^{ij}_{m_\alpha n_\beta}$ simply reads:

\begin{align}
\label{eq:KwithL}
K^{ij}_{m_\alpha n_\beta}=S_{m_\alpha i_\gamma}L^{ij}_{\gamma\delta}S_{j_\delta n_\beta}
.\end{align}
After applying the local excitation approximation,
this bi-electronic integral becomes even simpler, according to the following expression:

\begin{align}
\b{K}^{ij}=\b{S}^{ii}\b{L}^{ij}\b{S}^{jj}
.\end{align}

In the case of direct RPA, only the $\b{K}^{ij}$ two-electron integrals are needed. For the more general exchange RPA (RPAx) case, most of
the electron repulsion  integrals, $\braket{m_\alpha j}{n_\beta i}_\text{lr}$, can be neglected in the multipole approximation, since they correspond to the interaction of overlap charge densities formed by localized orbitals in  different domains. Nevertheless, integrals of the type
$\braket{m_\alpha j}{m_\beta j} _\text{lr}$ should be kept: they describe the interaction of the overlap charge densities of the $j$-th LMO and the $ m_\alpha$ POO, which is a typical long-range Coulomb interaction.
Similar considerations hold for the $\b{K}^\prime{}^{ij}$ matrix elements, which correspond to an exchange integral involving overlap charge densities of orbitals belonging to different domains and can be neglected at this point (a few integrals will however survive). The RPAx
variant of the model will be considered in more details in forthcoming works.

\subsection{Spherical average approximation}

The  previously discussed $3\times 3$
matrices can be easily replaced by scalar quantities, if one considers
a spherical average of the  POO overlap and fock matrices:

\begin{align}
  S^{ii}_{\alpha\beta}  \approx \tfrac{1}{3}\, s^i\,\delta_{\alpha\beta}
 \quad\text{with}\quad
  s^i = \sum_\alpha S^{ii}_{\alpha\alpha}
  \label{eq:siop}
,\end{align}
and:

\begin{align}
  f^{ii}_{\alpha\beta}  \approx \tfrac{1}{3}\, f^i\,\delta_{\alpha\beta}
 \quad\text{with}\quad
  f^i = \sum_\alpha f^{ii}_{\alpha\alpha}
.\end{align}

In this diagonal approximation, and in the case of direct RPA where $\b{A}=\b{B}=\b{K}$, the Riccati equations of Eq.~\ref{eq:RiccatiLocal} supposing implicit summations on $m$ and $n$ become:

\begin{align}
\label{eq:RiccatiDiagonal}
\b{R}^{ij}
&= s^i\,s^j\,\b{L}^{ij}
 + \left(f^i\,s^j
 - f_{ii}\,s^i\,s^j\right)\,\b{T}^{ij}
 + \tfrac{1}{3}\,s^i\,s^m\,s^j\,\b{L}^{im}\,\b{T}^{mj}
\nonumber\\&\qquad\qquad\,
 + \left(s^i\,f^j
 - s^i\,s^j\,f_{jj}\right)\,\b{T}^{ij}
 + \tfrac{1}{3}\,s^i\,s^m\,s^j\,\b{T}^{im}\,\b{L}^{mj}
\nonumber\\&\qquad\qquad\,
 + \tfrac{1}{3^2}\,s^i\,s^m\,s^n\,s^j\,\b{T}^{im}\,\b{L}^{mn}\,\b{T}^{nj}
=\b{0}
.\end{align}
This set of equations can be solved directly, \textit{i.e.}\ without proceeding by the pseudo-canonical transformation described in Appendix~\ref{app:WorkRic} for the more general case. The only quantities needed are  the spherically averaged  $s^i$ and $f^i$ associated to localized orbitals and the long-range dipole-dipole tensors. The update formula to get the $n$-th approximation to the amplitude matrix element is

\begin{align}
\label{eq:Tijiter}
 T^{ij~(n)}_{\alpha\beta} =
\frac{s^i\,s^j\,L^{ij}_{\alpha\beta}  +  \Delta R^{ij}_{\alpha\beta}(\b{T}^{(n-1)})}
     {\Delta^i\, s^j + s^i\, \Delta^j}
,\end{align}
with:

\begin{align}
\label{eq:Deltai}
 \Delta^i = f_{ii}\, s^i - f^i
,\end{align}
and:

\begin{align}
\Delta \b{R}^{ij}(\b{T})
&= \tfrac{1}{3} s^i\,s^m\,s^j\,\b{L}^{im}\,\b{T}^{mj}
  +\tfrac{1}{3} s^i\,s^m\,s^j\,\b{T}^{im}\,\b{L}^{mj}
\nonumber\\&\quad
  +\tfrac{1}{3^2}\, s^i\,s^m\,s^n\,s^j\,\b{T}^{im}\,\b{L}^{mn}\,\b{T}^{nj}
.\end{align}

Pursuing with the  local excitation and the spherical average approximations, the long-range correlation energy is given by the following spin-adapted expression:

\begin{align}
  E_\text{c}^{\text{RPA,lr}} = \frac{4}{9}
  \sum_{ij}^\text{occ}\, s^i\, s^j\,
  \text{tr}\,\bigl\{\b{L}^{ij}\,\b{T}^{ij}\bigr\}
.\end{align}

\subsection{Bond-bond C$_6$ coefficients}

Using the first order amplitudes, \textit{i.e.}\ the amplitudes obtained in the first iteration step during the solution of Eq.~\ref{eq:Tijiter}:

\begin{align}
  \b{T}^{{ij}~(1)}=
  \frac{s^i\,s^j}
       {\Delta^i\,s^j + s^i\,\Delta^j}\,\b{L}^{ij}
,\end{align}
the second order long-range correlation energy  becomes:

\begin{align}
  E_\text{c}^{\text{(2),lr}}  =
  \frac{4}{9}\,
  \sum_{ij}^\text{occ}\,
  \frac{s^i\,s^j\,s^i\,s^j}
       {\Delta^i\,s^j + s^i\,\Delta^j}
  \text{tr}\,\bigl\{\b{L}^{ij}\,\b{L}^{ij}\bigr\}
.\end{align}
This expression describes the correlation energy as a pairwise additive quantity made up from bond-bond contributions.
The summation over the components of the long-range interaction tensors gives (see Appendix~\ref{app:LRT}):

\begin{align}
  \text{tr}\,\bigl\{\b{L}^{ij}\,\b{L}^{ij}\bigr\}
  = \frac{6}{\b{D}^{ij}{}^6} F_\text{damp}^\mu(\b{D}^{ij})
,\end{align}
and allows us to cast the long-range correlation energy in a familiar form, as:

\begin{align}
  E_\text{c}^{\text{(2),lr}}
&=
 \sum_{ij}^\text{occ}\,
  \frac{\text{C}^{ij}_6}{{\b{D}^{ij}}^6}\,
       F_\text{damp}^\mu(\b{D}^{ij})
,\end{align}
where $\text{C}^{ij}_6$ is the dispersion coefficient between the $i$ and $j$ LMOs:

\begin{align}
\label{eq:C6rccd}
  \text{C}^{ij}_6 =
  \frac{8}{3}\,
  \frac{s^i\,s^j\,s^i\,s^j}
       {\Delta^i\,s^j + s^i\,\Delta^j}
.\end{align}
Note that the above dispersion coefficient corresponds to a single-term approximation to the  bare (non-interacting) spherically averaged dipolar dynamic polarizability associated with the localized orbital $i$,

\begin{align}
\label{eq:alphadyn}
  \over{\alpha}_0^i(i\omega) \approx
   \frac{4}{3}\,
  \frac{\over{\omega}_i}
       {\over{\omega}_i^2 + \omega^2}s^i
,\end{align}
where $\over{\omega}_i = \Delta^i/s^i$ is an effective energy denominator and the quantity $s^i$ stands for the second cumulant moment (spread) of the localized orbital.
Note that this possibility to approximate $\over{\alpha}_0^i(i\omega)$ only as a function of objects like $f^i$ and $s^i$ is a direct consequence of the remarkable feature that the second moment between an LMO and a POO corresponds to an overlap between two POOs (see Eq.~\ref{eq:remarkable}).

It is easy to verify that with the help of the Casimir-Polder formula

\begin{align}
  \text{C}^{ij}_6 = \frac{3}{\pi}\int_0^\infty \!\! d\omega\,\,
  \over{\alpha}_0^i(i\omega)\,\over{\alpha}_0^j(i\omega)
,\end{align}
that it is indeed the C$_6^{ij}$ dispersion coefficient which is recovered from the polarizabilities defined in Eq.~\ref{eq:alphadyn}. This simple model for the dynamic polarizability associated to an LMO, deduced from first principles, can be the starting point of alternative dispersion energy expressions, \textit{e.g.}\ based on the modeling of the dielectric matrix of the system or using the plasmonic energy expression. 

\section{Preliminary results: molecular C$_6$ coefficients}

In order to have a broad idea about the appropriateness of this simple dispersion energy correction, we have calculated the molecular C$_6$ coefficients for a series of homodimers as the sum of the atom-atom dispersion coefficients given by Eq.~\ref{eq:C6rccd}.

The matrix elements between POOs, $S^{ii}_{\alpha\beta}$ and $f^{ii}_{\alpha\beta}$, which are needed to calculate the scalars $s^i$ and $f^i$,
can be obtained directly by manipulating the matrix representation of the operators. Such a procedure
leads to what we will call the ``matrix algebra'' expressions (denoted by [M]), of the following form:

\begin{align}
\label{eq:siM}
& s^i_\text{[M]}
= \sum_\alpha \bra{i}\hat{r}_\alpha\,\hat{Q} \,\hat{r}_\alpha\ket{i}
= \sum_a^{\text{virt}}\, \vert\bra{i}\hat{\b{r}}\ket{a}\vert^2
,\end{align}
and (see Appendix~\ref{app:fockian}):

\begin{align}
\label{eq:fiM}
& f^i_\text{[M]}
= \sum_{ab}^{\text{virt}}\sum_\alpha\,
  \bra{i}\hat{r}_\alpha\ket{a}\,
  f_{ab}\,
  \bra{b}\hat{r}_\alpha\ket{i}
.\end{align}

Alternatively, these matrix elements can be obtained through the application of commutator relationships, therefore this latter option will be referred to as the ``operator algebra'' approach (denoted by [O]).
They take the form:

\begin{align}
\label{eq:siO}
  s^i_\text{[O]} 
= \sum_\alpha \bra{i}\hat{r}_\alpha\,\hat{Q} \,\hat{r}_\alpha\ket{i}
= \bra{i}\hat{\b{r}}^2\ket{i}
 -\sum_m^{\text{occ}}\, \vert\bra{i}\hat{\b{r}}\ket{m}\vert^2
,\end{align}
and (again, see Appendix~\ref{app:fockian}):

\begin{align}
\label{eq:fiO}
  f^i_\text{[O]}  = \tfrac{3}{2} & + \tfrac{1}{2}\sum_m^{\text{occ}}\,\bigl(
  f_{im}\bra{m}\hat{\b{r}}^2\ket{i} +
  \bra{i}\hat{\b{r}}^2\ket{m} f_{mi}
  \bigr)
\nonumber\\&\;\;\;
 -\sum_{mn}^{\text{occ}}\sum_\alpha\,
  \bra{i}\hat{r}_\alpha\ket{m}\,
  f_{mn}\,
  \bra{n}\hat{r}_\alpha\ket{i}
.\end{align}

Four different Fock/Kohn-Sham operators have been applied to obtain the orbitals, which are subsequently localized by the standard Foster-Boys procedure. In addition to the local/semi-local functionals LDA and PBE, the range-separated hybrid RSHLDA \cite{Angyan:05,Toulouse:09} with a range-separation parameter of $\mu=0.5$ a.u.\ as well as the standard restricted Hartree-Fock (RSH) method were used. The notations LDA[M] and LDA[O] refer to the procedure applied to obtain the matrix elements: either by the matrix algebra [M] or by the operator algebra [O] method. All calculations were performed with the aug-cc-pVTZ basis set, using the MOLPRO quantum chemical program package \cite{Molproshort:10}. The matrix elements were obtained by the MATROP facility of MOLPRO \cite{Molproshort:10}; the C$_6$ coefficients were calculated by Mathematica.

\begin{figure}
\begin{center}
\includegraphics[trim=26mm 0mm 7mm 0mm,clip=true,keepaspectratio=true,width=\linewidth]{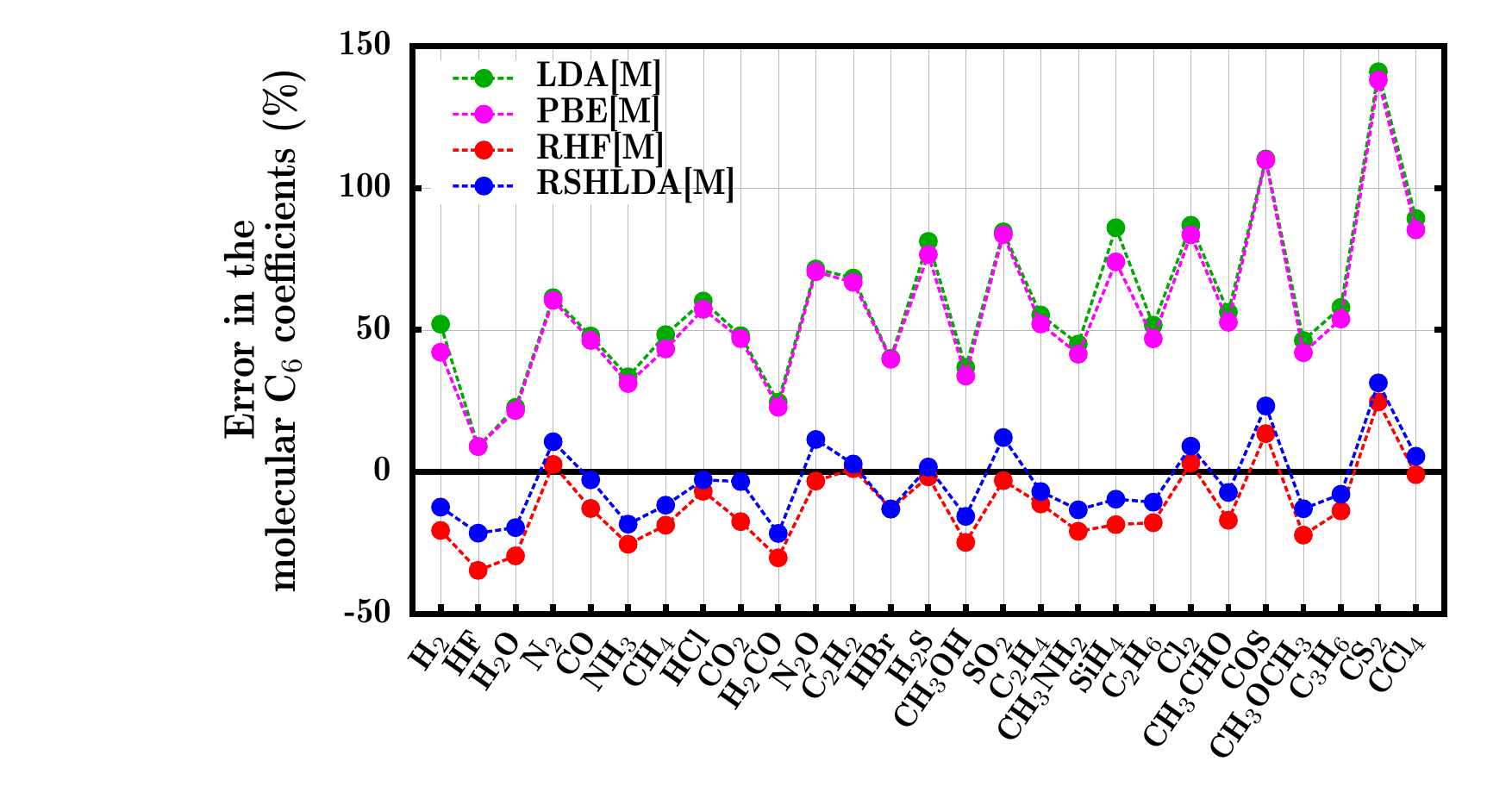}
\caption{Percentage errors in calculated molecular C$_6$ coefficients obtained with different methods using the matrix algebra approach.}
\label{fig:errors}
\end{center}
\end{figure}

\begin{table*}
\caption{Molecular C$_6$ coefficients from the dipolar oscillator orbital method, using LDA, PBE, RHF and sr-LDA/lr-RHF determinants in aug-cc-pVTZ basis set and Boys localized orbitals. The matrix elements are calculated with a matrix algebra [M] and operator algebra [O] approach, respectively. All results are in atomic units. Values highlighted as bold indicate the best agreement with experiment..}
\label{tab:allresults}
\begin{center}
\begin{tabular}{lrrrrrrrrr}
\hline\noalign{\smallskip}
Molecule& Ref.& \quad LDA[M] \quad    &  \quad LDA[O] \quad   & \quad  PBE[M]\quad     &  \quad PBE[O] \quad    & \quad RHF[M] \quad & \quad RHF[O]\quad   & RSHLDA[M]   &  RSHLDA[O] \\
\noalign{\smallskip}\hline\noalign{\smallskip}
\ce{H2}	& 12.1		& 18.4		& 18.4		& 17.2		& 17.2		& 9.6		& 16.2		& {\bf10.6}	& 16.9     \\
\ce{HF}	& 19.0		& 20.7		& 21.5		& 20.7		& 21.6		& 12.4		& 17.2		& 14.9		& {\bf19.3}\\
\ce{H2O}	& 45.3		& 55.6		& 56.5		& 55.1		& 56.1		& 31.9		& {\bf45.4}	& 36.4		& 48.9     \\
\ce{N2}	& 73.3		& 118.3		& 118.8		& 117.6		& 118.0		& {\bf75.2}	& 109.0		& 81.1		& 110.9    \\
\ce{CO}	& 81.4		& 120.4		& 120.9		& 119.1		& 119.7		& 70.9		& 104.1		& {\bf79.1}	& 109.6    \\
\ce{NH3}	& 89.0		& 118.8		& 119.9		& 116.7		& 117.8		& 66.3		& {\bf98.6}	& 72.6		& 102.0    \\
\ce{CH4}	& 129.7		& 192.4		& 193.0		& 185.9		& 186.5		& 105.3		& 163.5		& {\bf114.5}	& 168.2    \\
\ce{HCl}	& 130.4		& 208.9		& 211.8		& 205.1		& 208.5		& 121.5		& 186.3		& {\bf126.7}	& 184.8    \\
\ce{CO2}	& 158.7		& 234.9		& 237.0		& 233.2		& 235.4		& 130.9		& 184.0		& {\bf153.3}	& 202.8    \\
\ce{H2CO}	& 165.2		& 205.7		& 207.2		& 202.8		& 204.3		& 115.2		& {\bf168.3}	& 129.4		& 178.4    \\
\ce{N2O}	& 184.9		& 317.1		& 319.3		& 315.3		& 317.5		& {\bf178.9}	& 252.3		& 206.0		& 274.7    \\
\ce{C2H2}	& 204.1		& 343.5		& 345.2		& 340.4		& 342.2		& {\bf206.4}	& 316.1		& 209.7		& 306.5    \\
\ce{HBr}	& 216.6		& 303.3		& 325.9		& 302.5		& 325.4		& {\bf188.3}	& 293.5		& 188.2		& 282.7    \\
\ce{H2S}	& 216.8		& 392.8		& 397.7		& 382.8		& 388.2		& {\bf213.0}	& 339.2		& 220.5		& 335.5    \\
\ce{CH3OH}	& 222.0		& 303.8		& 305.3		& 297.0		& 298.5		& 166.8		& {\bf246.2}	& 187.2		& 260.2    \\
\ce{SO2}	& 294.0		& 542.6		& 554.9		& 539.8		& 552.5		& {\bf284.9}	& 416.7		& 329.5		& 456.2    \\
\ce{C2H4}	& 300.2		& 466.1		& 467.8		& 456.7		& 458.4		& 266.3		& 406.4		& {\bf279.4}	& 405.8    \\
\ce{CH3NH2}	& 303.8		& 440.6		& 442.3		& 429.8		& 431.4		& 240.2		& 360.3		& {\bf263.2}	& 373.3    \\
\ce{SiH4}	& 343.9		& 639.6		& 655.3		& 598.4		& 613.9		& 280.0		& 484.6		& {\bf310.9}	& 513.0    \\
\ce{C2H6}	& 381.9		& 579.2		& 580.9		& 560.9		& 562.5		& 313.5		& 480.9		& {\bf341.2}	& 495.5    \\
\ce{Cl2}	& 389.2		& 727.4		& 735.4		& 714.3		& 723.7		& {\bf401.3}	& 607.5		& 424.3		& 612.2    \\
\ce{CH3CHO}	& 401.7		& 627.5		& 630.4		& 613.3		& 616.0		& 333.2		& 493.6		& {\bf372.6}	& 521.7    \\
\ce{COS}	& 402.2		& 845.7		& 855.8		& 843.9		& 853.7		& {\bf456.3}	& 689.2		& 495.7		& 713.5    \\
\ce{CH3OCH3}	& 534.1		& 781.4		& 784.3		& 758.4		& 761.1		& 415.2		& 619.4		& {\bf464.8}	& 654.3    \\
\ce{C3H6}	& 662.1		& 1045.9	& 1049.4	& 1018.6	& 1021.8	& 571.2		& 868.2		& {\bf609.8}	& 881.2    \\
\ce{CS2}	& 871.1		& 2099.4	& 2119.5	& 2073.4	& 2094.5	& {\bf1085.6}	& 1658.0	& 1144.0	& 1686.2   \\
\ce{CCl4}	& 2024.1	& 3831.3	& 3861.0	& 3750.4	& 3784.0	& {\bf2004.8}	& 3007.0	& 2135.5	& 3051.9   \\
\noalign{\smallskip}\hline\noalign{\smallskip}
MAD\%E & & 59.84 &	61.69	&	56.71	&	58.62	&	15.22	&	33.75	&	11.84	 & 37.60 \\
STD\%E & & 28.08	  &	28.25	&	27.76	&	27.98	&	9.88	&	21.15	&7.24 &	20.86 \\
CSSD\%E && 67.14 &	68.92	&	64.11	&	65.97	&	18.39	&	40.37	&   14.07  & 	43.62 \\
\noalign{\smallskip}\hline\noalign{\smallskip}
\end{tabular}
\end{center}
\label{tab:default}
\end{table*}

\begin{table}
\caption{Detailed statistical analysis of the methods to obtain POO C$_6$ coefficients.}
\label{tab:stat_suggestion}
\centering
\begin{tabular}{lrrrr}
\hline\noalign{\smallskip}
	& LDA[M]& PBE[M]& RHF[M]& RSHLDA[M]\\
\hline\noalign{\smallskip}
MA\%E	& 59.8	& 56.7	& 15.2	& 11.8	\\
STD\%E	& 28.1	& 27.8	& 9.9	& 7.2	\\
CSSD\%E	& 67.1	& 64.1	& 18.4	& 14.1	\\
MED\%E	& 55.3	& 52.1	& 17.1	& 11.4	\\
MAX\%E	& 141	& 138	& 24.6	& 31.3	\\
MIN\%E	& 9.1	& 8.8	& -34.7	& -21.7	\\
\hline\noalign{\smallskip}
	& LDA2[M]& PBE2[M]& RHF2[M]& RSHLDA2[M]\\
\hline\noalign{\smallskip}
MA\%E	& 64.3	& 61	& 17.1	& 17.1	\\
STD\%E	& 35.7	& 35.3	& 12.3	& 12.3	\\
CSSD\%E	& 74.7	& 71.5	& 21.3	& 21.3	\\
MED\%E	& 53.5	& 50.6	& 15.3	& 15.3	\\
MAX\%E	& 142.1	& 148.7	& 54.3	& 54.3	\\
MIN\%E	& -4.7	& -5.1	& -36.6	& -36.6	\\
\hline\noalign{\smallskip}
	& LDA[O]& PBE[O]& RHF[O]& RSHLDA[O]\\
\hline\noalign{\smallskip}
MA\%E	& 61.7	& 58.6	& 33.7	& 37.6	\\
STD\%E	& 28.3	& 28	& 21.2	& 20.9	\\
CSSD\%E	& 68.9	& 66	& 40.4	& 43.6	\\
MED\%E	& 55.8	& 52.7	& 34	& 34.6	\\
MAX\%E	& 143.3	& 140.4	& 90.3	& 93.6	\\
MIN\%E	& 13.4	& 13.4	& -9.7	& 1.7	\\
\hline\noalign{\smallskip}
	& LDA2[O]& PBE2[O]& RHF2[O]& RSHLDA2[O]\\
\hline\noalign{\smallskip}
MA\%E	& 65.9	& 62.7	& 51	& 51	\\
STD\%E	& 36.3	& 36.9	& 47.7	& 47.7	\\
CSSD\%E	& 76.3	& 73.8	& 70.5	& 70.5	\\
MED\%E	& 54.7	& 51.1	& 34	& 34	\\
MAX\%E	& 143.3	& 152.9	& 214.8	& 214.8	\\
MIN\%E	& -1.1	& -1.2	& -10.6	& -10.6    \\
\hline\noalign{\smallskip}
\end{tabular}
\end{table}

The results and their statistical analysis are collected in Table~\ref{tab:allresults} for a set of small molecules taken from the database compiled by Tkatchenko and Scheffler \cite{Tkatchenko:09}, as used in \cite{Toulouse:13}. The experimental dispersion coefficients have been determined from dipole oscillator strength distributions (DOSD) \cite{Zeiss:77b,Margoliash:78,Kumar:08,Kumar:07,Kumar:05,Kumar:04,Kumar:03,Kumar:02a,Kumar:94,Kumar:92,Meath:90,Kumar:85b,Kumar:84}. The percentage errors of some of the methods (LDA[M], PBE[M], RHF[M] and RSHLDA[M]) are shown in Figure~\ref{fig:errors}.

Besides  Eq.~\ref{eq:C6rccd}, the C$_6$ coefficients were also calculated from an iterative procedure, where the amplitude matrices $\b{T}^{ij}$ were updated according to the first two lines of Eq.~\ref{eq:RiccatiDiagonal}. Such a procedure corresponds roughly to a local MP2 iteration, and the methods are labelled LDA2, PBE2, RHF2 and RSHLDA2. These results are summarized in Table~\ref{tab:stat_suggestion}, where a detailed statistical analysis is presented for all the computational results.

The dispersion coefficients obtained from LDA and PBE orbitals are strongly overestimated. It is not really surprising in view of the rather diffuse nature of DFA orbitals and their tendency to  underestimate the occupied/virtual gap. Due to the fact that LDA and PBE lead to local potentials, the operator and matrix algebra results are very similar: their difference is smaller than the supposed experimental uncertainty of the DOSD dispersion coefficients (few percent). It is quite clear from Figure~\ref{fig:errors} that the errors for the molecules containing second-row elements are considerably higher than for most of the molecules composed of H, C, N, O and F atoms. Notable exceptions are the  \ce{N2} and \ce{N2O} molecules.

The performance of the method is significantly better for orbitals obtained by nonlocal exchange: RHF and RSHLDA. In the latter case the long-range exchange is nonlocal, and only the short-range exchange is described by a short-range functional. The best performance has been achieved by the RSHLDA[M] method. In contrast to the pure DFA calculations, the matrix and operator algebra methods differ for the RHF and RSH methods significantly: the mean absolute error and the standard deviation of the error is increased by a factor of 2 or 3. This deterioration of the quality of the results reflects the fact that in the operator algebra approach the commutator of the position and non-local exchange operators is neglected (see Appendix~\ref{app:fockian}), while this effect is automatically taken into account in the matrix algebra calculations. In view of the simplicity of the model and of its fully ab initio character, the best ME\%E of about 11\% seems to be very promising and indicates that further work in this direction is worthwhile.

\section{Conclusions, perspectives}

It has been shown that, using projected dipolar oscillator orbitals to represent  the virtual space in a localized orbital context, the equations involved in long-range ring coupled cluster doubles type RPA calculations can be formulated without explicit knowledge of the virtual orbital set. The POO virtuals have been constructed directly from the localized occupied orbitals. The  fockian matrix elements and the electron repulsion integrals were evaluated using only the elements of the occupied block of the Fock/Kohn-Sham matrix in LMO basis and from simple multipole integrals between occupied LMOs. An interesting feature of the model is that, as far as one uses Boys localized orbitals and first-order dipolar oscillator orbitals, practically all the emerging quantities can be expressed by the overlap integral between two oscillator orbitals, which happens to be the matrix element of the dipole moment fluctuation operator. Various levels of approximations have been considered for the long-range RPA energy leading finally to a pairwise additive dispersion energy formula with a non-emprirical expression for the bond-bond C$_6$ dispersion coefficient.  At this simplest level a straightforward relationship has been unraveled between the ring coupled cluster and the dielectric matrix formulations of the long-range RPA correlation energy.

Our derivation, starting from the quantum chemical RPA correlation energy  
and pursuing a hierarchy of simplifying assumptions, has led us to a dispersion energy expression which is in a
straightforward analogy with classical van der Waals energy formulae. 
Our procedure produces an explicit model, derived from the wave function of the system, for the dynamical polarizabilities associated with the building blocks, which are bonds, lone pairs and in general localized electron pairs. Thus one arrives to the quantum chemical analogs of the ``quantum harmonic oscillators'' (QHO) appearing in the semiclassical theory of dispersion forces, elaborated within a RPA framework by Tkatchenko, Ambrosetti, di Stasio and their coworkers \cite{Tkatchenko:13a,Ambrosetti:14d}. 

A promising perspective of the projected oscillator orbital approach 
concerns the description of dispersion forces in plane wave calculations for solids. Much effort has been spent recently in finding a compact representation of the dielectric matrix in plane wave calculations, \cite{Rocca:14} but it still remains a bottleneck for a really fast non-empirical calculation of the long-range correlation energy. Projected oscillator orbitals may offer an opportunity to construct extremely compact representations of a part of the conduction band, which contributes the most to the local dipolar excitations at the origin of van der Waals forces. However, this approach is probably limited to relatively large gap semiconductors, where the construction of the maximally localized Wannier functions (MLWF), \cite{Marzari:97,Marzari:12a} which are the solid-state analogs of the Boys localized orbitals, is a convergent procedure. Such a methodology could become a fully non-empirical variant of Silvestrelli's
extension of the Tkatchenko-Scheffler approach for MLWFs~\cite{Silvestrelli:09,Silvestrelli:09a,Ambrosetti:12,Silvestrelli:13a,Silvestrelli:14}.  

The main purpose of the present work has been to describe the principal features of the formalism, without extensive numerical applications. However, the first rudimentary numerical tests on the molecular C$_6$ coefficients have indicated that the results are quite plausible in spite of the simplifying approximations and it is reasonable to expect that more sophisticated variants of the method  will improve the quality of the model. In view of the modest computational costs and the fully ab initio character of the projected oscillator orbital approach applied at various approximation levels of the RPA, which  is able to describe dispersion forces even beyond a pairwise additive scheme, the full numerical implementation of the presently outlined methodology have certainly a great potential for the treatment of London dispersion forces in the context of density functional theory.

\appendix

\section{Dipolar oscillator orbitals in local frame}
\label{app:POOlocframe}

Let $\pmb{\Ra}^i$ be the rotation matrix which transforms an arbitrary vector $\b{v}$ from the laboratory frame
to the vector $\b{v}^\text{loc}$ in the local frame defined as the principal axes of the second moment tensor of the charge distribution associated with a given localized occupied orbital $i$,
$\pmb{\Ra}^i\cdot \b{v} = \b{v}^\text{loc}$.
The expression in the local frame of a POO $i_\alpha$ constructed from the LMO $i$ is then:

\begin{align} 
   \ket{i_\alpha^\text{loc}} & =
 \left(\hat{I}-\sum_m^\text{occ.}\ket{m}\bra{m}\right) \bigl(\pmb{\Ra}^i\cdot (\b{r}-\b{D}^i)\bigr)_\alpha \ket{i}
   \nonumber \\
   & = \bigl(\pmb{\Ra}^i\cdot \b{r}\bigr)_\alpha \ket{i}
   - \sum_{m}^\text{occ.}
   \bigr(\pmb{\Ra}^i\cdot \bra{m}\b{r}\ket{i}\bigr)_\alpha \ket{m}
   \nonumber \\
   & = \sum_\beta \Ra^i_{\alpha\beta} r_\beta \ket{i}
   - \sum_{m}^\text{occ.}
   \sum_\beta \Ra^i_{\alpha\beta} \bra{m}r_\beta\ket{i} \ket{m}
\end{align}

\section{Riccati equations in POO basis}
\label{app:ricPOO}

The first order wave function $\Psi^{(1)}$ can be written in terms of
Slater determinants $\vert\ldots a        b      \ldots\vert$
formed with LMOs and canonical virtual orbitals $a$ and $b$
on the one hand, and on the other hand in terms of
Slater determinants $\vert\ldots m_\alpha n_\beta\ldots\vert$
formed with LMOs and POOs $m_\alpha$ and $n_\beta$.
That is to say that:

\begin{align}
  \Psi^{(1)}
 = \sum_{ij}^\text{occ}\,\sum_{ab}^\text{virt}\,
   T_{ab}^{ij} \vert\ldots  a        b     \ldots\vert
\approx
   \sum_{ij}^\text{occ}\,\sum_{m_\alpha n_\beta}^\POO\,
   T_{m_\alpha n_\beta}^{ij}
                \vert\ldots m_\alpha n_\beta\ldots\vert
.\end{align}
The canonical virtual orbitals and the POOs in question are related by (see Eq.~\ref{eq:virt2poo}):

\begin{align}
\ket{m_\alpha}=\sum_a^\text{virt.} \ket{a} V_{a m_\alpha}
\quad\text{and:}\quad
\ket{n_\beta} =\sum_b^\text{virt.} \ket{b} V_{b n_\beta}
.\end{align}
This allows us to write:

\begin{align}
  \Psi^{(1)}
&= \sum_{ij}^\text{occ}\,\sum_{ab}^\text{virt}\,
   T_{ab}^{ij} \vert\ldots  a        b     \ldots\vert
\nonumber\\&
\approx
 \sum_{ij}^\text{occ}\,\sum_{m_\alpha n_\beta}^\POO\,
   T_{m_\alpha n_\beta}^{ij}
   \vert\ldots m_\alpha n_\beta\ldots\vert
\nonumber\\&
 = \sum_{ij}^\text{occ}\,\sum_{m_\alpha n_\beta}^\POO\,
   \sum_{ab}^\text{virt}\,
   V^{\vphantom{\dagger}}_{a\,m_\alpha}\,T_{m_\alpha n_\beta}^{ij} V^\dagger_{n_\beta\,b}
   \vert\ldots a         b     \ldots\vert
,\end{align}
and leads to the transformation rule between the amplitudes in the VMO and in the POO basis:

\begin{align}
\label{eq:virttoPOO}
 \b{T}^{ij} = \b{V}\,\b{T}_\POO^{ij}\,\b{V}^\dagger
.\end{align}

Multiplication of the Riccati equations of Eq.~\ref{eq:RiccatiMO} by $\b{V}^\dagger$ and $\b{V}$ from the left and from the right respectively, and expressing the amplitudes in POOs using Eq.~\ref{eq:virttoPOO} leads to:

\begin{align}
  \b{V}^\dagger\,\b{R}^{ij}\,\b{V}
&=\b{V}^\dagger\,\b{B}^{ij}\,\b{V}
+ \b{V}^\dagger\,(\pmb{\epsilon}+\b{A})^{im}\left(\b{V}\,\b{T}_\POO^{mj}\,\b{V}^\dagger\right)\b{V}
\nonumber\\&\qquad\qquad\;\;
+ \b{V}^\dagger\left(\b{V}\,\b{T}_\POO^{im}\,\b{V}^\dagger\right)(\pmb{\epsilon}+\b{A})^{mj}\,\b{V}
\nonumber\\&\qquad\qquad\;\;
+ \b{V}^\dagger\left(\b{V}\,\b{T}_\POO^{im}\,\b{V}^\dagger\right)\b{B}^{mn}\left(\b{V}\,\b{T}_\POO^{nj}\,\b{V}^\dagger\right)\b{V}=\b{0}
,\end{align}
where  implicit summation conventions are supposed on $m$ and $n$.
Recognizing the expression for the overlap matrix, $\b{S}_\POO=\b{V}^\dagger\b{V}$, we obtain:

\begin{align}
\b{R}^{ij}_\POO
&= \b{B}^{ij}_\POO
 + (\pmb{\epsilon}_\POO+\b{A}_\POO)^{im}\,\b{T}_\POO^{mj}\,\b{S}^{~}_\POO
\nonumber\\&\qquad\quad\,\,
 + \b{S}^{~}_\POO\,\b{T}_\POO^{im}\,(\pmb{\epsilon}_\POO+\b{A}_\POO)^{mj}
\nonumber\\&\qquad\quad\,\,
 + \b{S}^{~}_\POO\,\b{T}_\POO^{im}\,\b{B}^{mn}_\POO\,\b{T}_\POO^{nj}\,\b{S}^{~}_\POO=\b{0}
,\end{align}
which defines $\b{R}^{ij}_\POO$, $\pmb{\epsilon}^{ij}_\POO$, $\b{A}^{ij}_\POO$ and $\b{B}^{ij}_\POO$ and which are the Riccati equations seen in Eq.~\ref{eq:RiccatiPOO}.

\section{Solution of the Riccati equations in POO basis}
\label{app:WorkRic}

To derive the iterative resolution of the Riccati equations seen in Eq.~\ref{eq:RiccatiPOO}, we write explicitly the fock matrix contributions hidden in the matrix $\pmb{\epsilon}$. The matrix elements in canonical virtual orbitals, $\epsilon^{ij}_{ab}$, read:

\begin{align}
\epsilon^{ij}_{ab}=\delta_{ij}f_{ab}-\delta_{ab}f_{ij}
,\end{align}
so that the matrix element in POOs is (we omit the ``POO'' indices):

\begin{align}
\epsilon^{ij}_{m_\alpha n_\beta}
=V^\dagger_{m_\alpha a}\epsilon^{ij}_{ab}V_{b n_\beta}
=\delta_{ij}f_{m_\alpha n_\beta} - S_{m_\alpha n_\beta}f_{ij}
.\end{align}
The terms in the Riccati equations containing the matrix $\pmb{\epsilon}$ then read (we use implicit summations over $m$ and $n$):

\begin{align}
\pmb{\epsilon}^{im}\,\b{T}^{mj}\,\b{S}
&=
 \b{f}\,\b{T}^{ij}\b{S}
-f_{im}\,\b{S}\,\b{T}^{mj}\b{S}
\\
\b{S}\,\b{T}^{im}\,\pmb{\epsilon}^{mj}
&=
 \b{S}\,\b{T}^{ij}\b{f}
-\b{S}\,\b{T}^{im}\,\b{S}\, f_{mj}
.\end{align}
We this in mind, the Riccati equations of Eq.~\ref{eq:RiccatiPOO} yield:

\begin{align}
\label{eq:RiccatiFockExplicit}
 \b{R}^{ij}
&=
 \b{B}^{ij} +
 \bigl(\b{f}\ -f_{ii}\,\b{S}\bigr)\,
  \b{T}^{ij}\b{S} +
 \b{S}\,\b{T}^{ij}\,\bigl(\b{f} -\b{S}\, f_{jj}\bigr)
\nonumber\\&\quad\quad\;\,
 - \sum_{m\neq i}\,
   f_{im}\,\b{S}\,\b{T}^{mj}\b{S}
 - \sum_{m\neq j}\,
   \b{S}\,\b{T}^{im}\,\b{S}\, f_{mj}
\nonumber\\&\quad\quad\;\,
 + \b{A}^{im}\,\b{T}^{mj}\b{S}
 + \b{S}\,\b{T}^{im}\b{A}^{mj}
 + \b{S}\,\b{T}^{im}\b{B}^{mn}\b{T}^{nj}\b{S}=\b{0}
.\end{align}
Remember that the matrices are of dimension $N_\POO\!\times\! N_\POO$.
Due to the nonorthogonality of the POOs and the non diagonal
structure of the fock matrix, the usual simple updating scheme for the
solution of the Riccati equations should be modified in a similar fashion as in the local coupled cluster theory \cite{Knowles:00}.
The fock matrix in the basis of the POOs will be diagonalized by the matrix $\b{X}$ obtained from  the solution  of the generalized eigenvalue problem:

\begin{align}
\label{eq:GenEigenEq}
\b{f}\,\b{X} = \b{S}\,\b{X}\,\pmb{\varepsilon}
.\end{align}
Note that the  transformation $\b{X}^\dagger\,\b{f}\,\b{X}$ does not brings us back to the canonical virtual orbitals.
We can write the transformation by the orthogonal matrix $\b{X}$ as
$X^\dagger_{\over{\vphantom{b}a}\,i_\alpha}\,f_{i_\alpha j_\beta}\,X_{j_\beta\,\over{b}}=\delta_{\over{\vphantom{b}a}\over{b}}\,\varepsilon_\over{b}$,
where $\over{\vphantom{b}a}$ and $\over{b}$ are
pseudo-canonical virtual orbitals that diagonalize the fock matrix expressed in POOs.
The Riccati equations of Eq.~\ref{eq:RiccatiFockExplicit}  are transformed separately for each pair $[ij]$ in the basis of the pseudo-canonical virtual orbitals that diagonalize $\b{f}_\POO$:

\begin{align}
  \b{X}^\dagger\,\b{R}^{ij}\,\b{X}
&=\b{X}^\dagger\,\b{B}^{ij}\,\b{X}
 +\bigl(\b{X}^\dagger\,\b{f} - f_{ii}\,\b{X}^\dagger\,\b{S}\bigr)\b{T}^{ij}\,\b{S}\,\b{X}
 +\b{X}^\dagger\,\b{S}\,\b{T}^{ij}\bigl(\b{f}\,\b{X} - \b{S}\,\b{X}\,f_{jj}\bigr)
\nonumber\\&\qquad\qquad\;\;
 -\sum_{m\neq i}\,f_{im}\,\b{X}^\dagger\,\b{S}\,\b{T}^{mj}\,\b{S}\,\b{X}
 -\sum_{m\neq j}\,\b{X}^\dagger\,\b{S}\,\b{T}^{im}\,\b{S}\,\b{X}\,f_{mj}
\nonumber\\&\qquad\qquad\;\;
 +\b{X}^\dagger\,\b{A}^{im}\,\b{T}^{mj}\,\b{S}\,\b{X}
 +\b{X}^\dagger\,\b{S}\,\b{T}^{im}\,\b{A}^{mj}\,\b{X}
\nonumber\\&\qquad\qquad\;\;
 +\b{X}^\dagger\,\b{S}\,\b{T}^{im}\,\b{B}^{mn}\,\b{T}^{nj}\,\b{S}\,\b{X}=\b{0}
,\end{align}
which can be simplified by the application of the generalized eigenvalue equation Eq.~\ref{eq:GenEigenEq}
and the use of the relationships $\b{I} = \b{S}\,\b{X}\,\b{X}^\dagger = \b{X}\,\b{X}^\dagger\,\b{S}$:

\begin{align}
\label{eq:RiccatiPseudo}
  \over{\b{R}}^{ij}
&=\over{\b{B}}^{ij} +
  \bigl(\pmb{\varepsilon} - f_{ii}\,\b{I}\bigr)\,\over{\b{T}}^{ij}
 +\over{\b{T}}^{ij}\bigl(\pmb{\varepsilon} - f_{jj}\,\b{I}\bigr)
\nonumber\\&\quad\quad\;\,
 - \sum_{m\neq i}\, f_{im}\,\over{\b{T}}^{mj}
 - \sum_{m\neq j}\, \over{\b{T}}^{im}f_{mj}
\nonumber\\&\quad\quad\;\,
 + \over{\b{A}}^{im}\,\over{\b{T}}^{mj}
 + \over{\b{T}}^{im}\,\over{\b{A}}^{mj}
 + \over{\b{T}}^{im}\,\over{\b{B}}^{mn}\,\over{\b{T}}^{nj}=\b{0}
,\end{align}
with the notations:

\begin{align}
\over{\b{R}}^{ij} & = \b{X}^\dagger\,\b{R}^{ij}\,\b{X}
\nonumber\\
\over{\b{A}}^{ij} & = \b{X}^\dagger\,\b{A}^{ij}\,\b{X}
\nonumber\\
\over{\b{B}}^{ij} & = \b{X}^\dagger\,\b{B}^{ij}\,\b{X}
\nonumber\\
\over{\b{T}}^{ij} & = \b{X}^\dagger\,\b{S}\,\b{T}^{ij}\,\b{S}\,\b{X}
\nonumber
.\end{align}

The new Riccati equations of Eq.~\ref{eq:RiccatiPseudo} can be solved by the iteration formula:

\begin{align}
  \over{T}^{ij~(n)}_{\over{\vphantom{b}a}\over{b}} = -
  \frac{\over{B}^{ij}_{\over{\vphantom{b}a}\over{b}} +
  \Delta\over{R}^{ij}_{\over{\vphantom{b}a}\over{b}}(\over{\b{T}}^{(n-1)})}
  {\varepsilon_\over{\vphantom{b}a} - f_{ii} +\varepsilon_\over{b} - f_{jj}}
,\end{align}
where $\Delta\over{\b{R}}^{ij}(\over{\b{T}})$ is

\begin{align}
  \Delta\over{\b{R}}^{ij}(\over{\b{T}})
&=-\sum_{m\neq i}\,f_{im}\,\over{\b{T}}^{mj}
  -\sum_{m\neq j}\,\over{\b{T}}^{im}f_{mj}
\nonumber\\&\quad
 +\over{\b{A}}^{im}\,\over{\b{T}}^{mj}
 +\over{\b{T}}^{im}\,\over{\b{A}}^{mj}
 +\over{\b{T}}^{im}\,\over{\b{B}}^{mn}\,\over{\b{T}}^{nj}
.\end{align}

As presented here, the update of the ``non-diagonal'' part of the residue is done in the pseudo-canonical basis. After convergence, we could transform the amplitudes back to the original POO basis according to:

\begin{align}
  \b{T}^{ij}
&= \bigl(\b{X}^\dagger\,\b{S}\bigr)^{-1}\,\over{\b{T}}^{ij}\,\bigl(\b{S}\,\b{X}\bigr)^{-1}
\nonumber\\&
 = \b{S}^{-1}(\b{X}^\dagger)^{-1}\,\over{\b{T}}^{ij}\,\b{X}^{-1}\,\b{S}^{-1}
\nonumber\\&
 = \b{X}\,\b{X}^\dagger(\b{X}^\dagger)^{-1}\,\over{\b{T}}^{ij}\,\b{X}^{-1}\,\b{X}\,\b{X}^\dagger
\nonumber\\&
 = \b{X}\,\over{\b{T}}^{ij}\,\b{X}^\dagger
.\end{align}
However, this back-transformation is not necessary since the correlation energy can be obtained directly in the pseudo-canonical basis, as:

\begin{align}
   \sum_{ij}^\text{occ}\, \text{tr}\bigl\{\over{\b{B}}^{ij}\,\over{\b{T}}^{ij} \bigr\}
&= \sum_{ij}^\text{occ}\, \text{tr}\bigl\{
 \b{X}^\dagger\,\b{B}^{ij}\,\b{X}\,\b{X}^\dagger\,\b{S}\,\b{T}^{ij}\,\b{S}\,\b{X}\bigr\}
\nonumber\\&
= \sum_{ij}^\text{occ}\, \text{tr}\bigl\{
 \b{B}^{ij}\,\b{T}^{ij}\,\b{S}\,\b{X}\,\b{X}^\dagger\bigr\}
=\sum_{ij}^\text{occ}\, \text{tr}\bigl\{
 \b{B}^{ij}\,\b{T}^{ij}\bigr\}
.\end{align}

\section{Riccati equations in the local excitation approximation}
\label{app:RiccLoc}

The local excitation approximation imposes that in the matrices $\b{R}$, $\pmb{\epsilon}$, $\b{A}$ and $\b{B}$, the excitations remain on the same localized orbitals.
In  this approximation the Riccati equations of Eq.~\ref{eq:RiccatiPOO},  with explicit virtual indexes, read:

\begin{align}
\label{eq:RiccExplIndex}
R^{ij}_{i_\alpha j_\beta}
&= B^{ij}_{i_\alpha j_\beta}
 + (\epsilon+A)^{im}_{i_\alpha m_\gamma}\,T^{mj}_{m_\gamma p_\delta}\,S^{~} _{p_\delta j_\beta}
\nonumber\\&\qquad\quad
 + S^{~} _{i_\alpha p_\gamma}\,T^{im}_{p_\gamma m_\delta}\,(\epsilon+A)^{mj}_{m_\delta j_\beta}
\nonumber\\&\qquad\quad
 + S^{~} _{i_\alpha p_\gamma}\,T^{im}_{p_\gamma m_\delta}\,B^{mn}_{m_\delta n_\tau}\,T^{nj}_{n_\tau q_\zeta}\,S^{~}_{q_\zeta j_\beta}=0
.\end{align}
In this context, the terms containing the matrix $\pmb{\epsilon}$ are (with explicit POO indexes):

\begin{align}
 \epsilon^{im}_{i_\alpha m_\gamma}\,T^{mj}_{m_\gamma p_\delta}\,S^{~} _{p_\delta j_\beta}
&=
 f_{i_\alpha i_\gamma}\,T^{ij}_{i_\gamma p_\delta}\, S_{p_\delta j_\beta}
-f_{im}\,S_{i_\alpha m_\gamma}\,T^{mj}_{m_\gamma p_\delta}\,S_{p_\delta j_\beta}
\\
 S^{~} _{i_\alpha p_\gamma}\,T^{im}_{p_\gamma m_\delta}\,\epsilon^{mj}_{m_\delta j_\beta}
&=
 S_{i_\alpha p_\gamma}\,T^{ij}_{p_\gamma j_\delta}\,f_{j_\delta j_\beta}
-S_{i_\alpha p_\gamma}\,T^{im}_{p_\gamma m_\delta}\,S_{m_\delta j_\beta}\, f_{mj}
.\end{align}
Inserting this in Eq.~\ref{eq:RiccExplIndex} and using the shorthand notation
$R^{ij}_{i_\alpha j_\beta}\equiv R^{ij}_{\alpha\beta}$,
$B^{ij}_{i_\alpha j_\beta}\equiv B^{ij}_{\alpha\beta}$,
$f     _{i_\alpha j_\beta}\equiv f^{ij}_{\alpha\beta}$ and
$S     _{i_\alpha j_\beta}\equiv S^{ij}_{\alpha\beta}$,
(note the matrix elements $T^{ij}_{i_\gamma p_\delta}$ cannot yet be translated to the shorthand notation) one obtains:

\begin{align}
R^{ij}_{\alpha \beta}
&= B^{ij}_{\alpha \beta}
+f^{ii}_{\alpha \gamma}\,T^{ij}_{i_\gamma p_\delta}\, S^{pj}_{\delta \beta}
-f_{im}\,S^{im}_{\alpha \gamma}\,T^{mj}_{m_\gamma p_\delta}\,S^{pj}_{\delta \beta}
+A^{im}_{\alpha \gamma}\,T^{mj}_{m_\gamma p_\delta}\,S^{pj}_{\delta \beta}
\nonumber\\&\quad\quad\;\;\,
+S^{ip}_{\alpha \gamma}\,T^{ij}_{p_\gamma j_\delta}\,f^{jj}_{\delta \beta}
-S^{ip}_{\alpha \gamma}\,T^{im}_{p_\gamma m_\delta}\,S^{mj}_{\delta \beta}\, f_{mj}
+S^{ip}_{\alpha \gamma}\,T^{im}_{p_\gamma m_\delta}\,A^{mj}_{\delta \beta}
\nonumber\\&\quad\quad\;\;\,
+S^{ip}_{\alpha \gamma}\,T^{im}_{p_\gamma m_\delta}\,B^{mn}_{\delta \tau}\,T^{nj}_{n_\tau q_\zeta}\,S^{qj}_{\zeta\beta}=0
.\end{align}

It is then a further approximation to tell that the POOs coming from different LMOs have a negligible overlap,
\textit{i.e.} that $S^{ij}_{\alpha\beta}=\delta_{ij} S^{ii}_{\alpha\beta}$. The Riccati equations become:

\begin{align}
R^{ij}_{\alpha \beta}
&= B^{ij}_{\alpha \beta}
+f^{ii}_{\alpha \gamma}\,T^{ij}_{i_\gamma j_\delta}\, S^{jj}_{\delta \beta}
-f_{ii}\,S^{ii}_{\alpha \gamma}\,T^{ij}_{i_\gamma j_\delta}\,S^{jj}_{\delta \beta}
+A^{im}_{\alpha \gamma}\,T^{mj}_{m_\gamma j_\delta}\,S^{jj}_{\delta \beta}
\nonumber\\&\quad\quad\;\;\,
+S^{ii}_{\alpha \gamma}\,T^{ij}_{i_\gamma j_\delta}\,f^{jj}_{\delta \beta}
-S^{ii}_{\alpha \gamma}\,T^{ij}_{i_\gamma j_\delta}\,S^{jj}_{\delta \beta}\, f_{jj}
+S^{ii}_{\alpha \gamma}\,T^{im}_{i_\gamma m_\delta}\,A^{mj}_{\delta \beta}
\nonumber\\&\quad\quad\;\;\,
+S^{ii}_{\alpha \gamma}\,T^{im}_{i_\gamma m_\delta}\,B^{mn}_{\delta \tau}\,T^{nj}_{n_\tau j_\zeta}\,S^{jj}_{\zeta \beta}=0
,\end{align}
which, in turn, allows us to use the shorthand notation 
$T_{i_\alpha j_\beta}\equiv T^{ij}_{\alpha\beta}$ to arrive at:

\begin{align}
\b{R}^{ij}
&= \b{B}^{ij}
+\b{f}^{ii}\,\b{T}^{ij}\,\b{S}^{jj}
-f_{ii}    \,\b{S}^{ii}\,\b{T}^{ij}\,\b{S}^{jj}
+\b{A}^{im}\,\b{T}^{mj}\,\b{S}^{jj}
\nonumber\\&\quad\quad\;\,
+\b{S}^{ii}\,\b{T}^{ij}\,\b{f}^{jj}
-\b{S}^{ii}\,\b{T}^{ij}\,\b{S}^{jj}\, f_{jj}
+\b{S}^{ii}\,\b{T}^{im}\,\b{A}^{mj}
\nonumber\\&\quad\quad\;\,
+\b{S}^{ii}\,\b{T}^{im}\,\b{B}^{mn}\,\b{T}^{nj}\,\b{S}^{jj}=0
,\end{align}

\section{Screened dipole interaction tensor}
\label{app:LRT}

Any interaction $L(\b{r})$ can be expanded in multipole series using a double Taylor expansion around appropriately selected centers, here $\b{D}^i$ and $\b{D}^j$, such that, with
$\b{r}=\b{r}^i-\b{r}^j=(\b{r}^i-\b{D}^i)+\b{D}^{ij}-(\b{r}^j-\b{D}^j)$ where $\b{D}^{ij}=\b{D}^i-\b{D}^j$:

\begin{align}
L(\b{r})
&= L^{ij}(\b{D}^{ij})
 +\sum_\alpha(\hat{r}^i_\alpha-D^i_\alpha)\,L^{ij}_\alpha(\b{D}^{ij})
 +\sum_\alpha(\hat{r}^j_\alpha-D^j_\alpha)\,L^{ij}_\alpha(\b{D}^{ij})
\nonumber\\&\quad
 +\sum_{\alpha\beta}(\hat{r}^i_\alpha-D^i_\alpha)(\hat{r}^j_\beta-D^j_\beta) L^{ij}_{\alpha\beta}(\b{D}^{ij}) +\ldots
,\end{align}
where the definitions of $L^{ij}_\alpha(\b{D}^{ij})$, $L^{ij}_{\alpha\beta}(\b{D}^{ij})$ are obvious.
For example, in the case of the long-range interaction, $L(\b{r})$ will be defined according to the RSH theory as:

\begin{align}
  L(\b{r}) =\frac{\text{erf}(\mu r)}{r}
,\end{align}
with $r=|\b{r}|$. The multipolar expansion of the long-range interaction leads to the following first and second order interaction tensors:

\begin{align}
  L^{ij}_\alpha(\b{D}^{ij}) & =
  - \frac{D^{ij}_\alpha}{{D^{ij}}^3}\,
  \biggl(1-\frac{2}{\sqrt{\pi}}\,{D^{ij}}\,\mu\,\text{e}^{-\mu^2{D^{ij}}^2}-\text{erf}(\mu {D^{ij}})\biggr)
   \\
  L^{ij}_{\alpha\beta}(\b{D}^{ij}) & =
  \frac{3\,D^{ij}_\alpha D^{ij}_\beta}{{D^{ij}}^5}\,
  \biggl(
  \text{erf}(\mu {D^{ij}})-
  \frac{2}{3\sqrt{\pi}}\,{D^{ij}}\,\mu\,
  \text{e}^{-\mu^2{D^{ij}}^2}\,\left(3+2 {D^{ij}}^2\mu^2\right)
  \biggr)
\nonumber\\&\quad
  -\frac{\delta_{\alpha\beta}{D^{ij}}^2}{{D^{ij}}^5}
  \biggl(
  \text{erf}(\mu {D^{ij}})-
  \frac{2}{\sqrt{\pi}}\,{D^{ij}}\,\mu\,
  \text{e}^{-\mu^2{D^{ij}}^2}
  \biggr)
.\end{align}
Remembering that the full-range Coulomb interaction tensor reads $T^{ij}_{\alpha\beta}(\b{D}^{ij})=\left(3\,D^{ij}_\alpha D^{ij}_\beta-\delta_{\alpha\beta}{D^{ij}}^2\right){D^{ij}}^{-5}$, the long-range interaction tensor can be written in an alternate form which clearly shows the damped dipole-dipole interaction contribution:

\begin{align}
 L^{ij}_{\alpha\beta}(\b{D}^{ij}) &=
 T^{ij}_{\alpha\beta}(\b{D}^{ij})\left(
 \text{erf}(\mu\,{D^{ij}})-
 \frac{2}{3 \sqrt{\pi}}\,{D^{ij}}\,\mu\,\text{e}^{-\mu^2 {D^{ij}}^2}\,
  \left(3+ 2 {D^{ij}}^2 \mu^2 \right)
   \right)
\nonumber\\&\quad
 -
  \delta _{\alpha \beta}\,e^{-\mu^2\, {D^{ij}}^2}\,
  \frac{4\,  \mu^3}{3 \sqrt{\pi}}
.\end{align}
The trace of the tensor product (used for the spherically averaged $C_6$) then reads:

\begin{align}
\sum_{\alpha\beta}\, L^{ij}_{\alpha\beta}L^{ij}_{\alpha\beta}
&=\frac{6}{{D^{ij}}^6}\,\biggl(
 4 \text{e}^{-2 {D^{ij}}^2 \mu ^2} {D^{ij}} \mu
 \biggl(
 \frac{ {D^{ij}} \mu \left(3+4 {D^{ij}}^2 \mu ^2+2 {D^{ij}}^4 \mu ^4\right)}{3 \pi}
\nonumber\\&\quad\quad\quad\quad\quad\quad\quad
-\frac{ \left(3+2 {D^{ij}}^2 \mu ^2\right)\text{erf}({D^{ij}} \mu)}{3 \sqrt{\pi}}
 \biggr)
+\text{erf}({D^{ij}} \mu)^2\biggr)
\nonumber\\&
=\frac{6}{{D^{ij}}^6}\,F_\text{damp}^\mu({D^{ij}})
.\end{align}

\section{Fock matrix element in POO basis}
\label{app:fockian}

The occupied-occupied block of the fock matrix, $f_{ij}$, is known.
The POOs are orthogonal to the occupied subspace of the original basis set, they satisfy the local Brillouin theorem, \textit{i.e.}\  the occupied-virtual block is zero.
As a result, in the local excitation approximation, we need only to deal with the fock matrix elements $f^{ii}_{\alpha\beta}$:

\begin{align}
f^{ii}_{\alpha\beta}=\bra{i_\alpha}\hat{f}\ket{i_\beta}
=\bra{i}\hat{r}_\alpha\hat{Q}\,\hat{f}\,\hat{Q}\,\hat{r}_\beta\ket{i}
,\end{align}
from which we directly derive the quantity $f^i_\text{[M]}$ of Eq.~\ref{eq:fiM}:

\begin{align}
f^i_\text{[M]}
&=\sum_\alpha f^{ii}_{\alpha\alpha}
 =\sum_{ab}^\text{virt} \bra{i}\hat{r}_\alpha\ket{a}\,f_{ab}\,\bra{b}\hat{r}_\alpha\ket{i}
.\end{align}

Since we would like to express everything in occupied orbitals, we expand the  projector $\hat{Q}$ and use that the occupied-virtual block of the fock matrix is zero to obtain the following expression:

\begin{align}
  f^{ii}_{\alpha\beta} =
  \bra{i}\hat{r}_\alpha\,\hat{f}\,\hat{r}_\beta\ket{i} -
  \sum_{mn}^\text{occ}\,
  \bra{i}\hat{r}_\alpha\ket{m}\,
  f_{mn}\,
  \bra{n}\hat{r}_\beta\ket{i}
.\end{align}
In order to transform the triple operator product, $\hat{r}_\alpha\,\hat{f}\,\hat{r}_\beta$, let us consider the following double commutator:

\begin{align}
\label{eq:doublecomm}
\bigl[\hat{r}_\alpha,[\hat{r}_\beta,\hat{f}]
\bigr] = - \delta_{\alpha\beta}
.\end{align}
Note that this holds provided that the fockian contains only local potential terms, which commute with the coordinate operator: see later for the more general case.
In this special case, the double commutator can be written as

\begin{align}
\bigl[\hat{r}_\alpha,[\hat{r}_\beta,\hat{f}]\bigr]
 = \hat{r}_\alpha\hat{r}_\beta\hat{f}
  -\hat{r}_\alpha \hat{f}\, \hat{r}_\beta
  -\hat{r}_\beta  \hat{f}\, \hat{r}_\alpha
  +\hat{f}\,\hat{r}_\beta\hat{r}_\alpha
=-\delta_{\alpha\beta}
,\end{align}
which allows us to express the two triple products:

\begin{align}
 \hat{r}_\alpha \hat{f}\, \hat{r}_\beta
+\hat{r}_\beta  \hat{f}\, \hat{r}_\alpha
 =  \delta_{\alpha\beta}
   +\hat{r}_\alpha\hat{r}_\beta \hat{f}
   +\hat{f}\,\hat{r}_\beta \hat{r}_\alpha
.\end{align}
The diagonal matrix element of the triple operator product is then:

\begin{align}
\bra{i}\hat{r}_\alpha \hat{f}\, \hat{r}_\beta\ket{i}
 =  \tfrac{1}{2}\delta_{\alpha\beta}
   +\tfrac{1}{2}\left(\bra{i}\hat{r}_\alpha\hat{r}_\beta\hat{f}\ket{i}
               + \bra{i}\hat{f}\,\hat{r}_\beta\hat{r}_\alpha\ket{i} \right)
.\end{align}
Since the localized orbitals satisfy local Brillouin theorem,
we finally obtain for the  matrix elements of the fock operator with multiplicative potential (typically Kohn-Sham operator with local or semi-local functionals) between two oscillator orbitals:

\begin{align}
  f^{ii}_{\alpha\beta}
= \tfrac{1}{2}\delta_{\alpha\beta}
&+ \tfrac{1}{2} \sum_m^\text{occ}\,
  \bigl(
  \bra{i}\hat{r}_\alpha\hat{r}_\beta\ket{m}f_{mi}
 +f_{im}\bra{m}\hat{r}_\alpha\hat{r}_\beta\ket{i}
  \bigr)
\nonumber\\&\quad
- \sum_{mn}^\text{occ}\,
  \bra{i}\hat{r}_\alpha\ket{m}\,
  f_{mn}\,
  \bra{n}\hat{r}_\beta\ket{i}
.\end{align}
From this, we obtain the quantity $f^i_\text{[O]}$ of Eq.~\ref{eq:fiO}:

\begin{align}
f^i_\text{[O]}=\sum_\alpha f^{ii}_{\alpha\alpha}
&=\tfrac{3}{2}
+\tfrac{1}{2} \sum_m^\text{occ}\,
  \bigl(
  \bra{i}\hat{\b{r}}^2\ket{m}f_{mi}
 +f_{im}\bra{m}\hat{\b{r}}^2\ket{i}
  \bigr)
\nonumber\\&\qquad\;\;
- \sum_{mn}^\text{occ}\,\sum_\alpha\,
  \bra{i}\hat{r}_\alpha\ket{m}\,
  f_{mn}\,
  \bra{n}\hat{r}_\alpha\ket{i}
.\end{align}

In the more general case, \textit{i.e.}\ when the fockian contains a nonlocal exchange operator, like in hybrid DFT and in Hartree-Fock calculations,
the relation seen Eq.~\ref{eq:doublecomm} does not hold any more and
the commutator of the position operator with the fockian contains an exchange contribution \cite{Starace:71,Harris:69},
which gives rise to an additional term:

\begin{align}
\bra{i} \bigl[\hat{r}_\alpha, [\hat{r}_\beta,\hat{K}]\bigr]\ket{i} =
  \sum_m^\text{occ}\,
 \bra{i m}
 \bigl(\hat{r}_\alpha-\hat{r}_\alpha^\prime\bigr)
  w(\b{r},\b{r}^\prime)
 \bigl(\hat{r}_\beta -\hat{r}_\beta^\prime \bigr)\ket{m i}
,\end{align}
where the nonlocal exchange operator is defined as

\begin{align}
\hat{K} = \sum_m^\text{occ}\,
          \int d\b{r}^\prime\,
          \phi_m^\dagger(\b{r}^\prime)\,
          w(\b{r},\b{r}^\prime)\,
          \hat{P}_{\b{r} \b{r}^\prime}\,
          \phi_m(\b{r}^\prime)
,\end{align}
where $\hat{P}_{\b{r} \b{r}^\prime}$ is the permutation operator that changes the coordinates $\b{r}^\prime$ appearing after $\hat{K}$ to $\b{r}$,
and we recall that $w(\b{r},\b{r}^\prime)$ is the two-electron interaction.
Hence, the diagonal blocks of the POO fockian in the general case can be written as:

\begin{align}
  \bra{i_\alpha}\hat{f}\ket{i_\beta}
&=\tfrac{1}{2}\,\delta_{\alpha\beta}
- \tfrac{1}{2}
  \sum_m^\text{occ}\,
 \bra{i m}
 \bigl(\hat{r}_\alpha-\hat{r}_\alpha^\prime\bigr)
  w(\b{r},\b{r}^\prime)
 \bigl(\hat{r}_\beta -\hat{r}_\beta^\prime \bigr)\ket{m i}
\nonumber\\&\qquad\quad\;
+ \tfrac{1}{2} \sum_m^\text{occ}\,
  \bigl(
  \bra{i}\hat{r}_\alpha\hat{r}_\beta\ket{m}f_{mi}
 +f_{im}\bra{m}\hat{r}_\alpha\hat{r}_\beta\ket{i}
  \bigr)
\nonumber\\&\qquad\quad\;
- \sum_{mn}^\text{occ}\,
  \bra{i}\hat{r}_\alpha\ket{m}\,
  f_{mn}\,
  \bra{n}\hat{r}_\beta\ket{i}
.\end{align}

In the present work, the exchange contribution,
which is present only in the case of Hartree-Fock or hybrid density functional fockians
and which would give rise to non-conventional two-electron integrals,
is not treated explicitly.
Possible approximate solutions for this problem will be discussed in forthcoming works.
Although we do not use the elements of the off-diagonal ($i \neq j$) blocks of the POO fock operator, for the sake of completeness we give its expression:

\begin{align}
  \bra{i_\alpha}\hat{f}\ket{j_\beta}
&=
- \bra{i} \hat{r}_\alpha \hat{\nabla}_\beta\ket{j}
  + \sum_m^\text{occ}\,
 \bra{i m}
 \hat{r}_\alpha\,
  w(\b{r},\b{r}^\prime)
 \bigl(\hat{r}_\beta -\hat{r}_\beta^\prime \bigr)\ket{m j}
\nonumber\\&\qquad\qquad\quad\;\;\;
+ \sum_m^\text{occ}\,
  \bra{i}\hat{r}_\alpha\hat{r}_\beta\ket{m}f_{mi}
\nonumber\\&\qquad\qquad\quad\;\;\;
- \sum_{mn}^\text{occ}\,
  \bra{i}\hat{r}_\alpha\ket{m}\,
  f_{mn}\,
  \bra{n}\hat{r}_\beta\ket{i}
.\end{align}

To derive this, instead of the double commutator of Eq.~\ref{eq:doublecomm}, one needs to consider the product of the commutator with a coordinate operator

\begin{align}
\hat{r}_\alpha [\hat{r}_\beta,\hat{f}] =
\hat{r}_\alpha [\hat{r}_\beta,\hat{T}] - \hat{r}_\alpha [\hat{r}_\beta,\hat{K}]
,\end{align}
where $\hat{T}$ is the kinetic energy operator, and $[\hat{r}_\beta,\hat{T}]=\hat{\nabla}_\beta$.

\begin{acknowledgements}
\normalsize
J.G.A.\ thanks Prof. P\'eter Surj\'an
(Budapest), to whom this article is dedicated,  the fruitful discussions
during an early stage of this work.
\end{acknowledgements}

\end{document}